\documentclass[aps,amsmath,notitlepage,twocolumn,amssymb,prb,longbibliography]{revtex4-1}
\usepackage{gpkmac}
\usepackage{graphicx}
\usepackage{wasysym}
\usepackage{stmaryrd}
\usepackage{verbatim}

\begin{document}

\title{Pyrochlore U(1) spin liquid of mixed symmetry enrichments in magnetic fields}
\author{Xu-Ping Yao$^{1}$}    
\author{Yao-Dong Li$^{2}$}    
\author{Gang Chen$^{1,3,4}$}  
\affiliation{$^{1}$Department of Physics and HKU-UCAS Joint Institute for
Theoretical and Computational Physics at Hong Kong, The University of Hong Kong, Hong Kong, China}
\affiliation{$^{2}$Department of Physics, University of California, Santa Barbara, California 93106, USA} 
\affiliation{$^{3}$State Key Laboratory of Surface Physics and Department of Physics, 
Fudan University, Shanghai 200433, China}
\affiliation{$^{4}$Collaborative Innovation Center of Advanced Microstructures, 
Nanjing University, Nanjing 210093, China}
\date{\today}

\begin{abstract}
We point out the experimental relevance and the detection scheme of symmetry 
enriched U(1) quantum spin liquids (QSLs) outside the perturbative spin-ice regime. 
Recent experiments on Ce-based pyrochlore QSL materials suggest that the candidate 
QSL may not be proximate to the well-known spin ice regime, and thus differs 
fundamentally from other pyrochlore QSL materials. We consider the possibility 
of the $\pi$-flux U(1) QSL favored by frustrated transverse exchange interactions 
rather than the usual quantum spin ice. It was previously suggested that both 
dipolar U(1) QSL and octupolar U(1) QSL can be realized for the generic spin model 
for the dipole-octupole doublets of the Ce$^{3+}$ local moments on the pyrochlore  
magnets Ce$_2$Sn$_2$O$_7$ and Ce$_2$Zr$_2$O$_7$. 
We explain and predict the experimental signatures especially the magnetic field 
response of the octupolar $\pi$-flux U(1) QSL. Fundamentally, this remarkable state 
is a mixture of symmetry enrichments from point group symmetry and from translational 
symmetry. We discuss the relevant experiments for pyrochlore U(1) QSLs and further 
provide some insights to the pyrochlore Heisenberg model. 
\end{abstract}

\maketitle

\section{Introduction}

Symmetry is the key that underlies the traditional Landau's paradigm of 
many-body phases and phase transitions. It is almost so in the classification 
and understanding of topological and exotic phases of quantum matter~\cite{Wen200710}. 
In last decade or so, a tremendous progress has been made theoretically 
to classify various symmetry enriched topological phases, 
where symmetry creates much more
topological phases~\cite{PhysRevB.65.165113,Hermele2013,Ran2013,Levin2005,
PhysRevB.99.085128,PhysRevB.94.195120,CHEN20173}. 
These symmetry enriched topological phases are described by the same 
topological quantum field theory, but they are distinct by the 
realization of symmetries for example on the fractionalized excitations. 
These beautiful theories so far do not have strong experimental connections. 
It is thus of a great interest to find an experimental relevance and 
establish the connection.

In last decade or so, various quantum spin liquid (QSL) candidate 
materials have been proposed, and the rare-earth pyrochlore magnets
comprise an important and large family of materials~\cite{PhysRevLett.98.157204,
Ross11,Shannon2012,PhysRevB.90.014429,PhysRevB.95.094407,PhysRevB.92.054432,
PhysRevLett.118.107206,PhysRevLett.109.097205,PhysRevLett.121.037203,
PhysRevB.94.104430,PhysRevLett.107.207207,Lhotel2014,Chang2014,Yasui2003,
Romain2017,Zhihao2014,PhysRevLett.108.247210,PhysRevLett.105.047201,
PhysRevB.96.125145,PhysRevB.86.054411,Udagawa} in these proposals. 
In these materials, the rare-earth ions carry spin-orbital-entangled effective 
spin-1/2 local moments that interact with highly anisotropic superexchange 
interactions~\cite{PhysRevB.78.094418,Ross11,PhysRevB.83.094411,Onoda_2011,huang1311quantum}
Due to the proximity to the classical spin ice regime where the classical 
Ising interaction dominates, many pyrochlore materials develop a spin ice 
type of Pauling entropy plateau at low but finite temperatures~\cite{Castelnovo12008,
castelnovo1112spinice, Gingras2001,gingras1311quantum,RevModPhys.82.53,
PhysRevLett.95.217201,PhysRevLett.115.037201,PhysRevLett.98.157204}. 
Introducing quantum fluctuations and/or perturbations to the extensively 
degenerate spin ice manifold could then convert the system into a QSL state, 
and this state is often quoted as quantum spin ice U(1) QSL or pyrochlore 
ice U(1) QSL~\cite{hermele0305photon,huse0305coulomb, motrunich0407origin,
PhysRevLett.98.157204,gingras1311quantum,Ross11,savary1110coulomb}.
Is the proximity to the spin ice regime necessary to produce a U(1) QSL?
In our opinion, this condition was merely a theoretical convenience 
to access the interesting and exotic state in early theoretical 
works~\cite{hermele0305photon,savary1110coulomb}.
It is now established that, the pyrochlore U(1) QSL is much more 
robust in the so-called frustrated regime where the spinon experiences 
an emergent background $\pi$ flux~\cite{lee1204generic,chen1704spectral,
PhysRevX.7.041057,PhysRevLett.121.067201}. Since this $\pi$-flux U(1) QSL 
is expected to extend much beyond the perturbative spin ice regime~\cite{lee1204generic}, 
it is natural to expect that the proximity to the spin ice regime 
is not quite necessary to obtain the pyrochlore U(1) QSL. 
We refer the U(1) QSL in this regime as non-spin-ice pyrochlore U(1) QSL 
or simply as pyrochlore U(1) QSL instead of pyrochlore spin ice U(1) QSL.

In the actual experiments on the Ce-based pyrochlore QSL materials 
(in particular, Ce$_2$Zr$_2$O$_7$)~\cite{dai1901experimental}, 
there does not exist the spin ice type of Pauling entropy plateau down to 
very low temperatures while the magnetic entropy is almost completely exhausted.
This is a clear indication that the system is not in the spin ice regime. 
Another interesting aspect is that the Ce$^{3+}$ local moment in both 
Ce$_2$Sn$_2$O$_7$~\cite{sibille1502candidate,2019arXiv191200928S,2020arXiv200110304L} 
and Ce$_2$Zr$_2$O$_7$ is 
a dipole-octupole doublet~\cite{huang1311quantum, li1607symmetry, li1608hidden}.
It is thus natural for us to consider the possibility of pyrochlore 
U(1) QSL beyond the spin ice regime with the dipole-octupole doublets. 
It was previously suggested that, the anisotropic interaction 
between the dipole-octupole doublets on the pyrochlore lattice 
could stabilize two symmetry enriched U(1) QSLs, i.e. 
dipolar U(1) QSL and octupolar U(1) QSL~\cite{huang1311quantum,li1607symmetry}. 
The major distinction between these two U(1) QSLs arises from the 
transformation of emergent electric field under the point group 
symmetry operation, i.e., the 
emergent electric field in the dipolar U(1) QSL transforms as the
magnetic dipole moment while
the emergent electric field in the octupolar U(1) QSL transforms as 
the magnetic octupole moment. 
On top of this point group symmetry enrichment, there is an additional 
translational symmetry enrichment where the spinon could experience
a background $\pi$ flux or $0$ flux in the distinct enrichments. 
It was shown that,~\cite{lee1204generic} 
the $\pi$ flux state (labelled as U(1)$_{\pi}$ QSL) extends much 
beyond the perturbative ice regime. Therefore, it is reasonable to 
associate the non-spin-ice pyrochlore QSL with the U(1)$_{\pi}$ QSL. 
The 0 flux state (labelled as U(1)$_{0}$ QSL) has been studied
extensively in the previous literature~\cite{savary1110coulomb,lee1204generic,huang1311quantum}. 
For the dipole-octupole doublets, the octupolar U(1)$_{0}$ QSL 
has been studied by us in a previous work~\cite{li1607symmetry},
and in the current work, we will mostly focus on the octupolar U(1)$_{\pi}$ 
QSL and explore its physical properties.


The octupolar U(1)$_{\pi}$ QSL is the quantum phase that most clearly 
reflects the interplay between the multipolar nature of the local 
moments and emergent exotic properties of the U(1) QSL. 
The strong frustrated interaction between the octupolar components 
is the precondition for realizing the octupolar U(1)$_{\pi}$ QSL. 
In terms of the emergent degrees of freedom for the octupolar U(1)$_{\pi}$ 
QSL, the octupole component is the emergent electric field whose 
correlation contains both the gapess U(1) gauge photon and 
the gapped ``magnetic monopoles''. These magnetic octupole components, 
however, do not couple with the external magnetic field and the 
neutron spin at the linear order. Thus, they are hidden from the 
conventional measurements. What is visible is the spinon sector. 
The external magnetic field couples linearly with the dipole 
component that does not commute with the octupole component or 
the emergent electric field. Thus it is observed that~\cite{li1607symmetry}, 
the external magnetic field couples with the spinon-antispinon pair
and modifies the spinon dispersion. For the octupolar U(1)$_{\pi}$ QSL,
the spinon continuum has a spectral periodicity enhancement due to the 
background $\pi$ flux. We specifically study the experimental signatures 
of the octupolar U(1)$_{\pi}$ QSL for the frustrated non-spin-ice regime
and explore the spinon continuum and 
the magnetic excitations under the magnetic fields.

The remaining parts of the paper are organized as follows. 
In Sec.~\ref{sec2}, we introduce the model for the dipole-octupole doublets 
on the pyrochlore lattice and emphasize the unique coupling to magnetic fields. 
In Sec.~\ref{sec3}, we explain the connection between the microscopic degrees 
of freedom and the emergent degrees of freedom in the octupolar U(1)$_{\pi}$ QSL.
In Sec.~\ref{sec4}, we explore the impact of the external magnetic field on 
the spinon continuum in the octupolar U(1)$_{\pi}$ QSL.
In Sec.~\ref{sec5}, we analyze the spin-wave spectrum in the 
regime with strong magnetic fields. Finally in Sec.~\ref{sec6}, 
we discuss some experimental relevance and the related theoretical questions.

\section{Effective spin model}
\label{sec2}

We start with the generic effective spin model for the dipole-octupole doublets 
on the pyrochlore lattice. The model was derived in Ref.~\onlinecite{huang1311quantum}
and is known as the XYZ 
model~\cite{huang1311quantum, li1607symmetry, li1608hidden},
\begin{eqnarray}
H &=& \sum_{\langle ij \rangle} 
J_z S^z_i S^z_j + J_x S^x_i S^x_j + J_y S^y_i S^y_j \nonumber \\
&&      + J_{xz} (S^x_i S^z_j + S^z_i S^x_j) - h \sum_i 
(\hat{n}\cdot \hat{z}_i ) S_i^z ,
\label{eq1}
\end{eqnarray} 
where {\sl microscopically} $S^x$ and $S^y$ are magnetic octupole moments 
while $S^z$ is a magnetic dipole moment. From the symmetry analysis, 
$S^x$ and $S^z$ transform identically under the point group symmetry. Thus, 
$S^x$ is sometimes referred as the magnetic dipole moment~\cite{huang1311quantum}.
We have also introduced the 
Zeeman coupling that only acts on the magnetic dipole moment $S^z$, 
and $\hat{n}$ is the field direction and $\hat{z}_i$ defines the 
local $z$ direction of each sublattice (see Appendix~\ref{appA} for 
definition of these conventions). Only the nearest-neighbor interaction 
is considered here, which is expected to be reasonable for the localized 
$4f$ electrons. 
The XYZ form is obtained by applying a rotation around the $y$-direction 
by an angle $\theta$ to eliminate the $J_{xz}$ term; the resulting Hamiltonian reads
\begin{eqnarray}
H_{\text{XYZ}} &=& \sum_{\langle ij \rangle} 
\tilde{J}_z \tilde{S}^z_i \tilde{S}^z_j + \tilde{J}_x \tilde{S}^x_i 
\tilde{S}^x_j + \tilde{J}_y \tilde{S}^y_i \tilde{S}^y_j              \nonumber \\
&&            - h \sum_i 
(\hat{n}\cdot \hat{z}_i ) ( \cos \theta \tilde{S}_i^z + \sin \theta \tilde{S}_i^x ),
\label{eq2}
\end{eqnarray}
where $\tilde{S}_i^{x, z}$ are related to ${S}_i^{x, z}$ by the $\theta$-rotation, 
and ${\tilde{S}^y \equiv S^y}$. In the phase diagram of $H_{\text{XYZ}}$ without
 the magnetic field, the system supports three disconnected U(1) 
 QSLs~\cite{li1607symmetry,huang1311quantum}. When $\tilde{J}_z$ ($\tilde{J}_y$) 
 is antiferromagnetic and dominant while the remaining two couplings 
 are not large enough to drive a magnetic order, the ground state is 
 a dipolar (an octupolar) U(1) QSL. In the case when $\tilde{J}_x$ is 
 antiferromagnetic and large, the relevant U(1) QSL is regarded as 
a dipolar U(1) QSL and shares the 
same universal and qualitatively similar physics with the dipolar 
U(1) QSL because $\tilde{S}^x$ and $\tilde{S}^z$ transform identically 
under the point group symmetry. The dipolar U(1) QSL and the octupolar 
U(1) QSL are symmetry enriched U(1) QSLs and are enriched by the point group symmetry.

\begin{figure*}[t]
	\includegraphics[width=.3\textwidth]{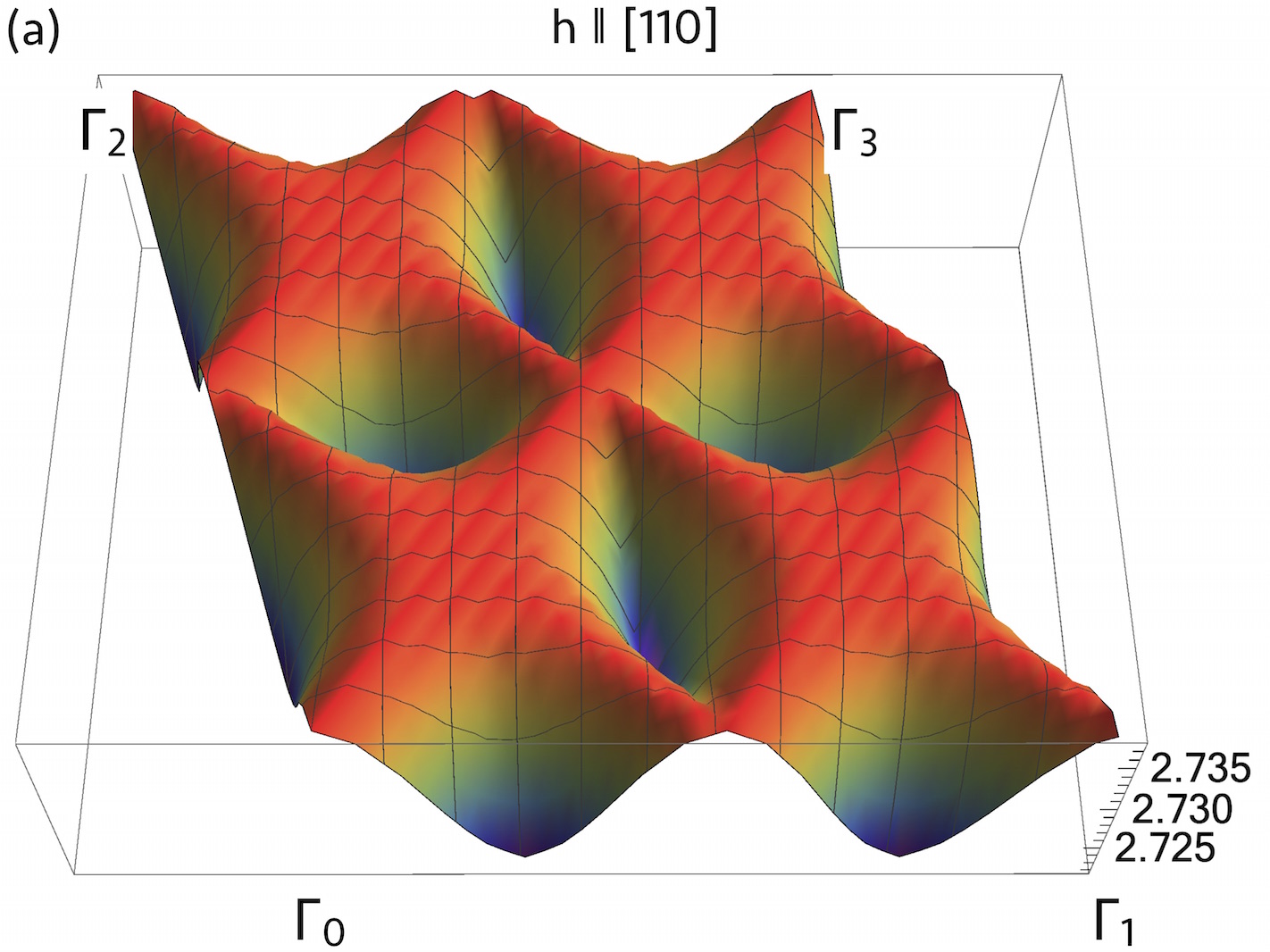}
    \includegraphics[width=.3\textwidth]{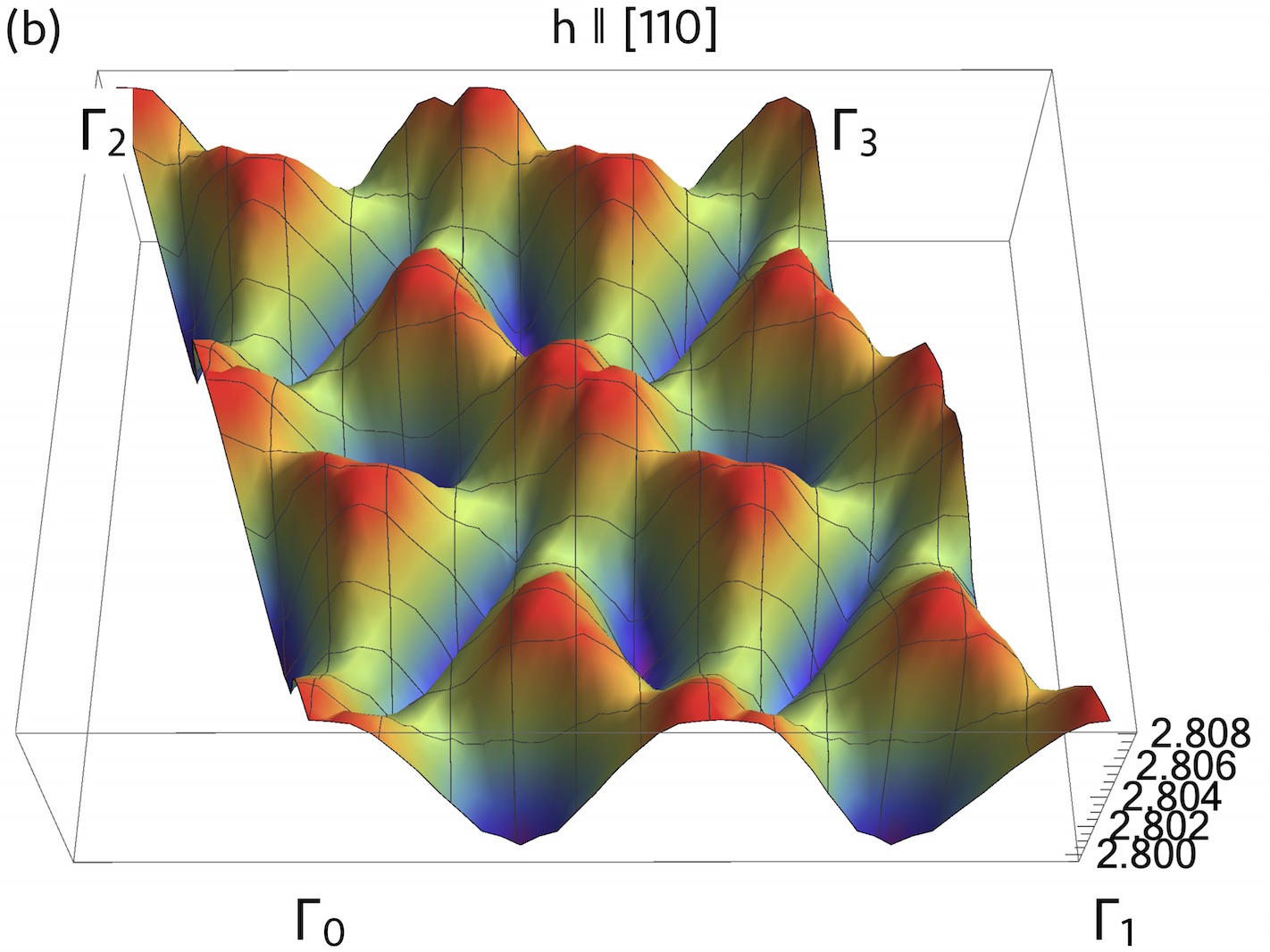}
    \includegraphics[width=.3\textwidth]{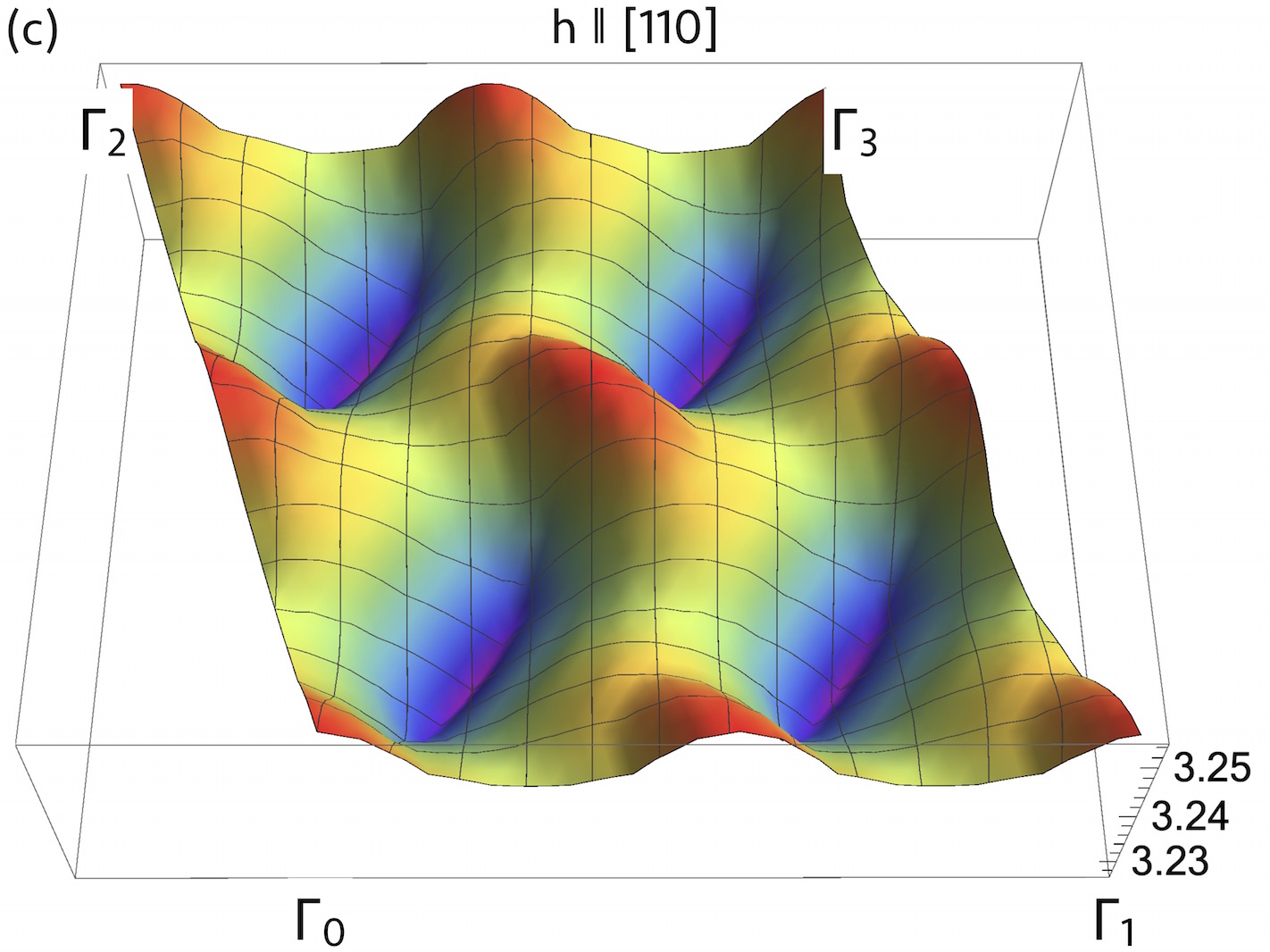}
	\includegraphics[width=.3\textwidth]{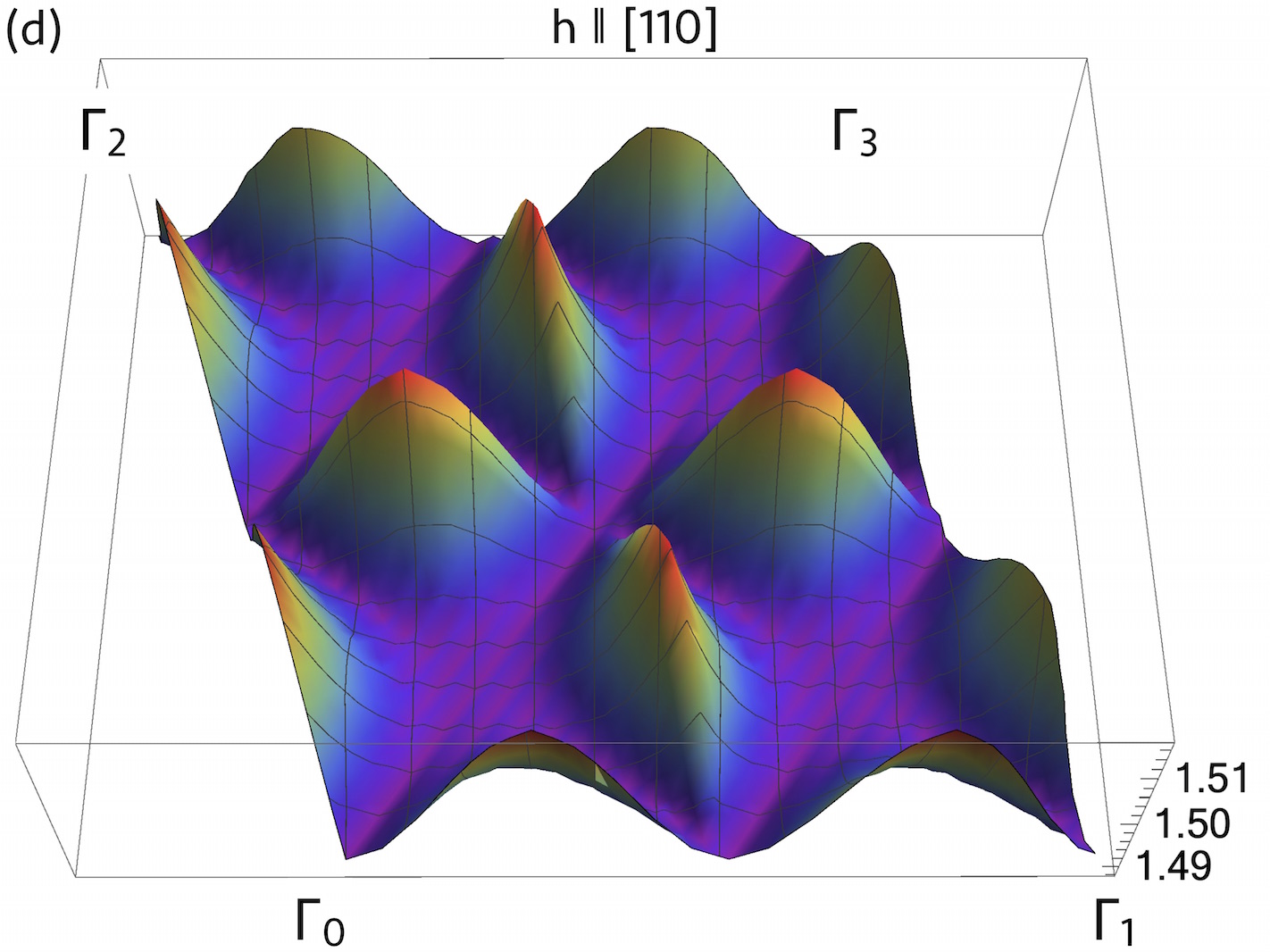}
    \includegraphics[width=.3\textwidth]{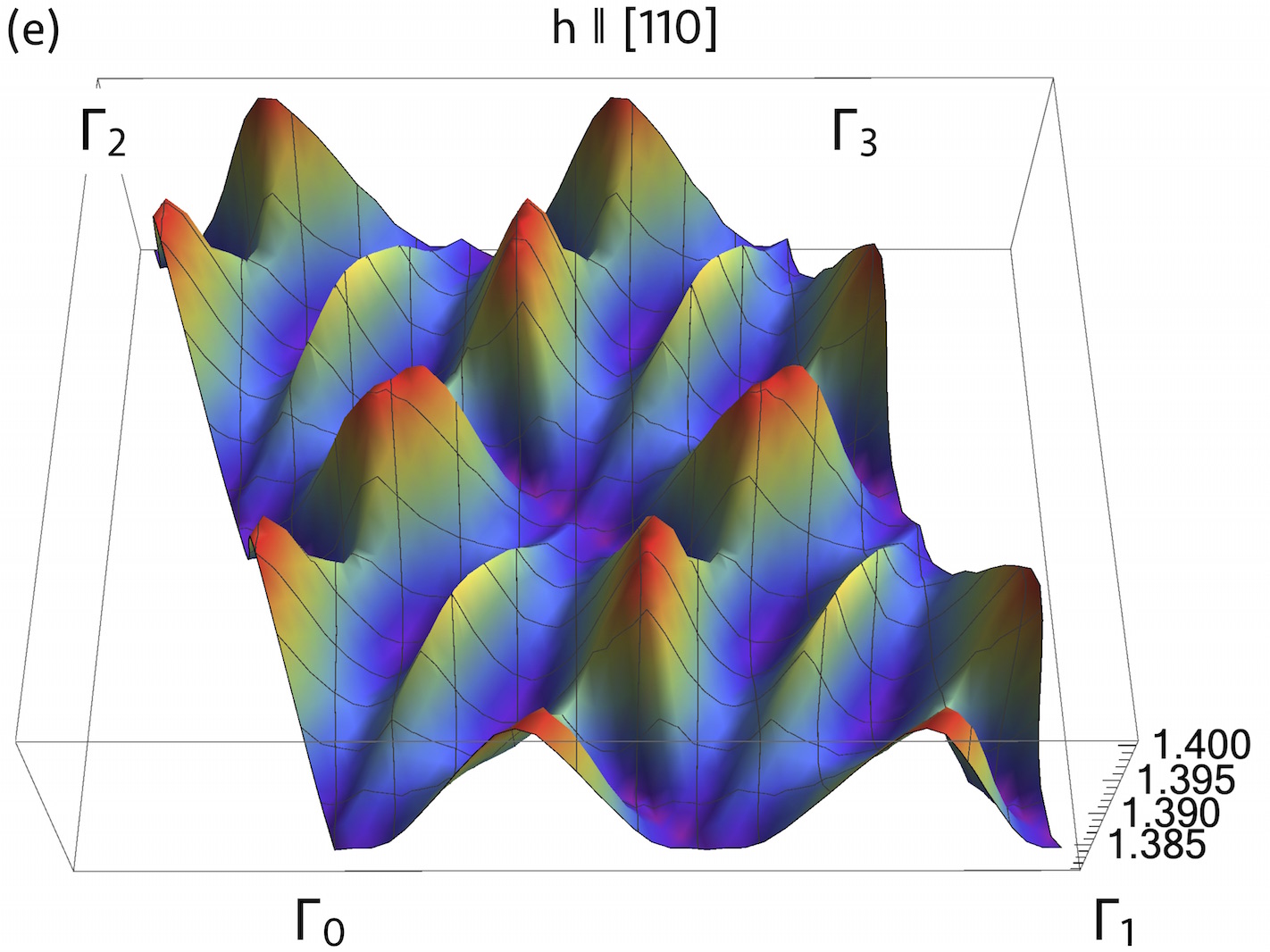}
    \includegraphics[width=.3\textwidth]{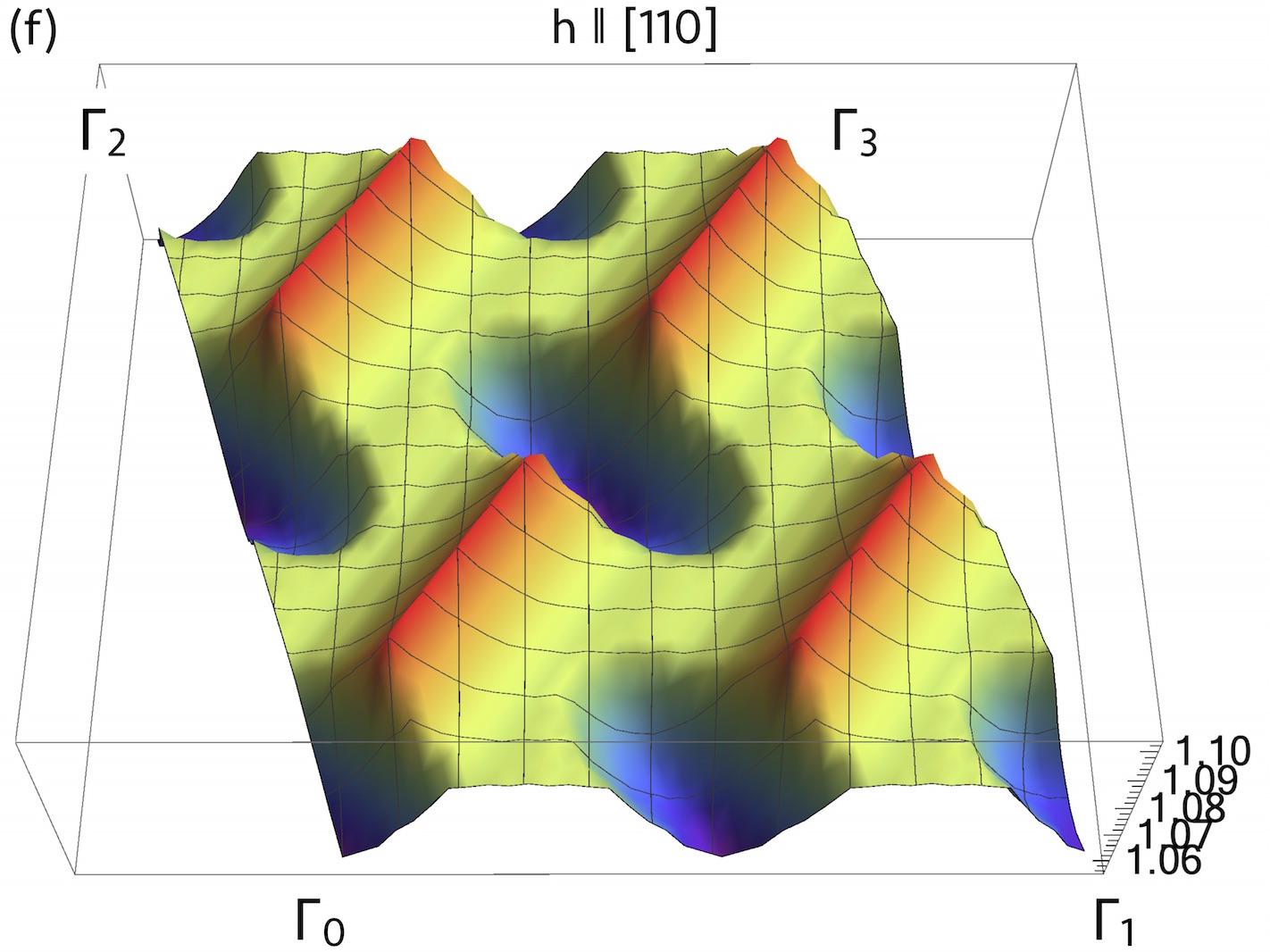}
\caption{(Color online.) The upper and lower excitation edges of 
the two-spinon continuum in the octupolar U(1)$_\pi$ QSL, 
for the magnetic fields along the [110] direction.
We fix ${J_{\pm}^y = -0.33 J_y},{J_{\pm\pm}^y}=0, {\theta=0}$, 
and take ${h = 0}$ in (a, d), ${h = 0.2 J_y}$ in (b, e), 
and ${h = 1.0 J_y}$ in (c, f). The energy unit is set to $J_y$.
Here, the $\Gamma$ points represent the Brillouin zone centers and 
differ by the reciprocal lattice vectors with ${\Gamma_0\Gamma_1= 2\pi(-1,1,1)}$ 
and ${\Gamma_0\Gamma_2  = 2\pi(1,-1,1)}$.}
    \label{fig:excitation_edge}
\end{figure*}

\section{Octupolar U(1)$_\pi$ QSL}
\label{sec3}

Since the experiments suggest that Ce$_2$Zr$_2$O$_7$ 
is not in the spin ice regime~\cite{dai1901experimental}, 
we would like to understand this from the physical properties of 
both dipolar and octupolar U(1) QSLs in the non-spin-ice regime. 
From the previous 
argument and early results~\cite{lee1204generic}, the non-spin-ice regime for a XXZ model  
would be in the frustrated regime with a frustrated transverse exchange interaction
and support the U(1) QSL with $\pi$ flux for spinons. 

The XYZ model with zero magnetic field can be rewritten with two different 
but equivalent forms below,
\begin{widetext}
\begin{eqnarray}
\label{H_z}
 {H}_z & = & \sum_{\langle ij \rangle } 
\tilde{J_z} \sigma^z_i \sigma^z_j 
- J_{\pm}^z (\sigma^+_i \sigma^-_j + \sigma^-_i \sigma^+_j) 
  + J_{\pm\pm}^z (\sigma^+_i \sigma^+_j + \sigma^-_i \sigma^-_j) ,
  \\
  \label{H_y}
   {H}_y & = & \sum_{\langle ij \rangle } 
\tilde{J_y} \tau^y_i \tau^y_j 
- J_{\pm}^y ({\tau}^+_i  {\tau}^-_j +  {\tau}^-_i  {\tau}^+_j) 
  + J_{\pm\pm}^y ( {\tau}^+_i {\tau}^+_j +  {\tau}^-_i  {\tau}^-_j) ,
\end{eqnarray}
\end{widetext}
where we have 
\begin{eqnarray}
&&\sigma_i^z \equiv \tilde{S}_i^z, \quad {\sigma^\pm_i \equiv \tilde{S}^x_i \pm i \tilde{S}^y_i} \\
&&\tau^y \equiv \tilde{S}_i^y,     \quad {\tau^\pm_i \equiv \tilde{S}^z_i \pm i \tilde{S}^x_i}
\end{eqnarray} 
and the couplings $(J_{\pm}^z,J_{\pm\pm}^z)$ and $(J_{\pm}^y, J_{\pm\pm}^y)$ can be read 
off from the expansion the above two Hamiltonians into the original form. 
For convenience, we focus on the regime where the ground state of $H_z$ ($H_y$) 
is the dipolar (octupolar) U(1) QSL, i.e. when $\tilde{J}_z$ ($\tilde{J}_y$) is 
antiferromagnetic and dominant. It is known that, as long as ${J_\pm^z > 0}$ 
(${J_\pm^y > 0}$), the model for either sign of $J_{\pm\pm}^z$ ($J_{\pm\pm}^y$)
does not have a fermion sign problem for quantum Monte carlo simulation~\cite{huang1311quantum}. 
In this unfrustrated regime, numerics shows that the system has
the classical spin ice phenomena such as the Pauling entropy plateau 
at low and finite temperatures even when the system is located 
in the QSL phase at zero temperature~\cite{huang1806extended}.
It means that the frustrated regime ${J_\pm^z < 0}$ (${J_\pm^y < 0}$) 
should carry the QSL physics for the Ce-based pyrochlore magnets.
Since the frustrated regime for the U(1) QSL generates an emergent 
$\pi$-flux for the spinons, it is then natural to understand the 
physical properties of the dipolar and octupolar U(1)$_{\pi}$ QSLs. 
It is interesting to note that the $\pi$ flux for the spinons is 
a signature of the symmetry enrichments in the lattice translation of the spinon sector.
This is a translational symmetry enrichment on top of the point group symmetry enrichments.
Due to the $\pi$ flux, we expect the spinon continuum to develop an enhanced spectral periodicity 
in the reciprocal space with a folded Brillouin zone~\cite{chen1704spectral,PhysRevB.90.121102}.
Although certain generic properties may be established from the model level, 
there is still a gap to a quantitative connection to the actual 
physical observables of the dipole-octupole doublets.

To make connection with the experiments, it is important to notice that only $S^z$ 
in Eq.~\eqref{eq1} is magnetic~\cite{huang1311quantum, li1607symmetry, li1608hidden}, 
and only $S^z$-$S^z$ correlation is measurable in a neutron scattering experiment. 
From Eq.~\eqref{eq2}, ${S^z_i = \cos \theta \sigma_i^z + \sin \theta \sigma_i^x}$.
Thus, the inelastic neutron scattering experiment would measure both $\sigma^z$-$\sigma^z$ 
and $\sigma^x$-$\sigma^x$ correlators. For the dipolar U(1)$_{\pi}$ QSL of $H_z$ 
with a large and antiferromagnetic $\tilde{J}_z$ (for the dipolar U(1)$_{\pi}$ 
QSL with a large and antiferromagnetic $\tilde{J}_x$), the spinon continuum is 
contained in the $\sigma^x$-$\sigma^x$ ($\sigma^z$-$\sigma^z$) correlator, and 
the ``magnetic monopole'' continuum and the gauge photon are contained in the 
$\sigma^z$-$\sigma^z$ ($\sigma^x$-$\sigma^x$) correlator.
The inclusion of the ``magnetic monopole'' continuum was understood quite 
recently~\cite{chen1706dirac}. Due to the background $\pi$ flux for 
the spinons in the dipolar U(1)$_{\pi}$ QSL, the spinon continuum 
develops an enhanced spectral periodicity with a folded Brillouin 
zone~\cite{chen1706dirac, chen1704spectral,lee1204generic}. 
For the ``magnetic monopoles'', the continuum should always have an enhanced 
spectral periodicity with a folded Brillouin zone due to the effective 
spin-1/2 nature of the local moment~\cite{chen1706dirac,chen1704spectral,PhysRevB.94.205107}. 
As for the octupolar U(1)$_{\pi}$ QSL, because $\tau^y$ is not directly 
measurable, the ${S}^z$-${S}^z$ correlator only detects the gapped spinon 
continuum, and the continuum has an enhanced spectral 
periodicity~\cite{li1607symmetry, chen1706dirac, chen1704spectral}.
\\

\section{Evolution of spinon continuum under magnetic fields for octupolar U(1)$_{\pi}$ QSL}
\label{sec4}

To access the ground state and illustrate the emergent U(1) gauge 
structure and the physical properties of the XYZ spin model, 
we implement the mapping introduced in Refs.~\onlinecite{savary1110coulomb,lee1204generic} 
of the spin model to an Abelian-Higgs model with the compact U(1) gauge field 
and the bosonic spinon matter. Focusing on the octupolar U(1) QSL regime 
(when $\tilde{J}_y$ is positive and dominant in Eq.~\eqref{H_y}), 
we express the spin operators as 
\envelope{eqnarray}{  
    {\tau^{y}_{\bs{r}, \bs{r} + \bs{e}_\mu} = s^{y}_{\bs{r}, \bs{r} + \bs{e}_\mu} }, 
    \quad
    {\tau^{+}_{\bs{r}, \bs{r} + \bs{e}_\mu} = \Phi^\dg_{\bs{r}} s^{+}_{\bs{r}, \bs{r} + \bs{e}_\mu} 
    \Phi^\pg_{\bs{r} + \bs{e}_\mu}},
}
where $\bs{r}$ belongs to the I diamond sublattice 
(our convention is summarized in Appendix~\ref{appA}).
Here $s^y_{\bs{r}, \bs{r}^\p}$ is the emergent electric field in the octupolar U(1) QSL phase, 
$s^\pm_{\bs{r}, \bs{r}^\p}$ is the gauge string operator ending at sites $\bs{r}$ and 
$\bs{r}^\p$, and $\Phi^\pg_{\bs{r}}$ ($\Phi^\dg_{\bs{r}}$) 
is the spinon annihilation (creation) operator at the diamond lattice 
site $\bs{r}$. The physical Hilbert space is obtained by imposing the 
following constraints
\envelope{eqnarray}{
	Q_\bs{r} = \eta_\bs{r} \sum_\mu s^y_{\bs{r}, \bs{r} + \eta_\bs{r} \bs{e}_\mu},
        \quad
    \Phi^\dg_{\bs{r}} \Phi^\pg_{\bs{r}} = 1,
}
where ${\eta_{\bs{r}} = \pm 1}$ for $\bs{r}$ in sublattice I and II, respectively, 
and $Q_{\bs{r}}$ is the operator measuring the local gauge charge through the ``Gauss law'', 
and is canonically conjugate to $\Phi_{\bs{r}}$,
\envelope{eqnarray}{
	\lz \Phi^\pg_{\bs{r}}, Q_{\bs{r}} \rz = \Phi^\pg_{\bs{r}}, \quad
	\lz \Phi^\dg_{\bs{r}}, Q_{\bs{r}} \rz = \Phi^\dg_{\bs{r}}.
}
Under this mapping the Hamiltonian becomes
\begin{widetext}
\begin{eqnarray}
H_y &=& 
        \frac{\tilde{J}_y}{2} \sum_{\bs{r}} Q_\bs{r}^2
        - J_{\pm}^y \sum_\bs{r} \sum_{\mu \neq \nu} \(
            \Phi^\dg_{\bs{r} + \eta_\bs{r} \bs{e}_\mu}
            \Phi^\pg_{\bs{r} + \eta_\bs{r} \bs{e}_\nu}
            s^{-\eta_\bs{r}}_{\bs{r}, \bs{r} + \eta_\bs{r} \bs{e}_\mu}
            s^{+\eta_\bs{r}}_{\bs{r}, \bs{r} + \eta_\bs{r} \bs{e}_\nu}
        \)  \nn
        && + \frac{J_{\pm\pm}^y}{2} \sum_\bs{r} \sum_{\mu \neq \nu} \(
            \Phi^\dg_{\bs{r}}
            \Phi^\dg_{\bs{r}}
            \Phi^\pg_{\bs{r} +\eta_\bs{r} \bs{e}_\mu}
            \Phi^\pg_{\bs{r} +\eta_\bs{r} \bs{e}_\nu}
            s^{+\eta_\bs{r}}_{\bs{r}, \bs{r} + \eta_\bs{r} \bs{e}_\mu}
            s^{+\eta_\bs{r}}_{\bs{r}, \bs{r} + \eta_\bs{r} \bs{e}_\nu}
            + h.c.
        \) \nn
        && 
        - \frac{h}{4} \sum_\bs{r} \sum_\mu \(\hat{ {n}}\cdot \hat{  {z}}_{\bs{r}+\eta_\bs{r} \bs{e}_\mu / 2} \) \lz
          \cos\theta \( \Phi^\dg_{\bs{r}} \Phi^\pg_{\bs{r} + \eta_\bs{r} \bs{e}_\mu} 
                          s^{+\eta_\bs{r}}_{\bs{r}, \bs{r} + \eta_\bs{r} \bs{e}_\mu} + h.c. \)
            -i \sin \theta \( \Phi^\dg_{\bs{r}} \Phi^\pg_{\bs{r} + \eta_\bs{r} \bs{e}_\mu} 
                          s^{+\eta_\bs{r}}_{\bs{r}, \bs{r} + \eta_\bs{r} \bs{e}_\mu} - h.c. \)
        \rz.
\label{eqLGT}
\end{eqnarray}
\end{widetext}
Within the U(1)$_\pi$ QSL regime, we choose a gauge to take care of the 
background $\pi$-flux~\cite{lee1204generic, chen1704spectral}, such that 
the spinons hop on the diamond lattice with modulated signs of hoppings 
(see Appendix~\ref{appA}). In the absence of the field, the spinon continuum,
that is measurable via an inelastic neutron scattering measurement in the octupolar 
U(1)$_\pi$ QSL, shows a spectral periodicity enhancement with a folded Brillouin
zone. As we calculate explicitly and show in the left panel of Fig.~\ref{fig:excitation_edge},
both the upper and lower excitation edges of the two-spinon continuum develop
the spectral periodicity enhancement. Another advantage of the octupolar U(1) 
QSL is to allow the external magnetic field to tune the spinon dispersion 
directly even in the presence of the background $\pi$ flux.

The external magnetic field, that couples to $S^z$ or equivalently couples 
to the spinon matters, modifies the spinon band structures.
This modification can then be directly measured by the inelastic neutron scattering
probe. 
This provides an interesting example to manipulate or control the emergent fractionalized 
spinon degrees of freedom with external means that is the external magnetic field here. 
More importantly, such a manipulability could be recorded and tested experimentally.  
We apply the fields along three high symmetry directions, i.e. 
[001], [110] and [111] crystallographic directions. In 
the central panel and the right panel of Fig.~\ref{fig:excitation_edge}, 
we plot the upper and lower excitation edges of the spinon continuum under 
two different magnetic fields along the [110] direction. Because the 
weak magnetic field does not revise the background $\pi$ flux, 
the spinon continuum in these plots continues to 
develop an enhanced spectral periodicity with a folded Brillouin zone. 
This important topological property remains to be the distinct feature 
to be examined even in the presence of the magnetic field. 
The detailed calculation scheme and the results for the fields along 
the [001] and [111] directions are displayed in Appendix~\ref{appB} 
and Appendix~\ref{appC}. Despite the application of the magnetic fields,
the enhanced spectral periodicity preserves, and the magnetic field also 
generates non-universal features such as the rich wiggles in the spectra.

The above calculation is based on the assertion that the 3D
U(1) QSL is stable against the perturbation from the weak magnetic fields. 
What happens if the field becomes strong? To address this question, 
we notice that there is a (hidden) competition between 
the transverse spin exchange interaction and the magnetic field. 
Our observation is as follows. The strong magnetic field would simply
favor an uniform polarized state that preserves the lattice translations, 
while the simple spinon condensation of the U(1)$_\pi$ QSL would favor 
a state that breaks the lattice translational symmetry~\cite{chen1704spectral}. 
This frustration could enhance the stability of U(1)$_\pi$ QSL against 
the external magnetic field. The stability of U(1)$_\pi$ QSL against 
exchange interactions and other competing orders
has been previously established in Ref.~\onlinecite{lee1204generic} 
and Ref.~\onlinecite{PhysRevLett.121.067201}, respectively. 
This might also be the reason for the more stability of the antiferromagnetic 
Kitaev QSL in the magnetic field over the ferromagnetic one~\cite{PhysRevB.97.241110}.
Perturbatively, the magnetic field favors a zero-flux state. One 
may wonder if the field can drive a phase transition between 
two symmetry enriched U(1) QSLs, i.e., from U(1)$_\pi$ to U(1)$_0$ QSLs,
and then from U(1)$_0$ QSL to the spinon condensed state, or a direct 
first order transition from U(1)$_\pi$ QSL to the polarized state,
or the field first drives a spinon condensation by breaking the 
lattice translation and then restores the lattice transition by
entering a polarized phase via a first order transition. 
This may be examined numerically or experimentally.

\section{Magnetic excitations in the strong field regime}
\label{sec5}

As the external magnetic field is further increased, the system will eventually 
enter a polarized state. For the fully or nearly polarized state, the spins 
(or the local $z$ components) are aligned along the preferred direction 
according to the external magnetic field. Since the transverse spin 
components that create the coherent spin excitations are the octupolar moments, 
the neutron spin does not couple linearly with the transverse spin component 
and thus the inelastic neutron scattering signal would be suppressed. However, 
there can still be residual intensity for the nearly polarized state due to 
the crossing coupling $J_{xz} (S_i^x S_j^z + S_i^z S_j^x)$. This can be 
understood as follows. Although the magnetic field polarizes the $S^z$ 
components directly, the finite $S^z$ would further induce a finite 
$S^x$ through the crossing coupling. As a result, the $S^z$ operator 
could create coherent magnetic excitations by flipping $S^x$ components.   


The distinction between the dipolar U(1) QSL and the octupolar U(1) QSL 
not only appears in the qualitative behaviors under the weak magnetic fields 
or by the neutron scattering measurements, but also shows up in the magnetic 
excitations when the QSL state is replaced by the polarized states in the 
strong magnetic fields. The former has been explained in the previous sections. 
The latter can be simply understood as follows. We directly compare the
dipolar U(1) QSL with a dominant $\tilde{J}_z$ with the octupolar U(1) 
QSL with a dominant $\tilde{J}_y$. Regardless of which U(1) QSL 
the system is located in, it is always the transverse spin component
that flips the $S^z$ components and generates the spin wave excitation
in the polarized state. As the transverse couplings for two distinct 
U(1) QSLs are very different compared to the dominant interactions, 
it is meaningful to explore quantitatively the spin wave dispersion 
under different magnetic fields in different symmetry enriched U(1) 
QSLs with distinct parameter regimes, and this information would in 
principle be able to distinguish which U(1) QSL the polarized state 
may be originated from.

\begin{figure}[th]
    \centering
    \includegraphics[width=0.47\textwidth]{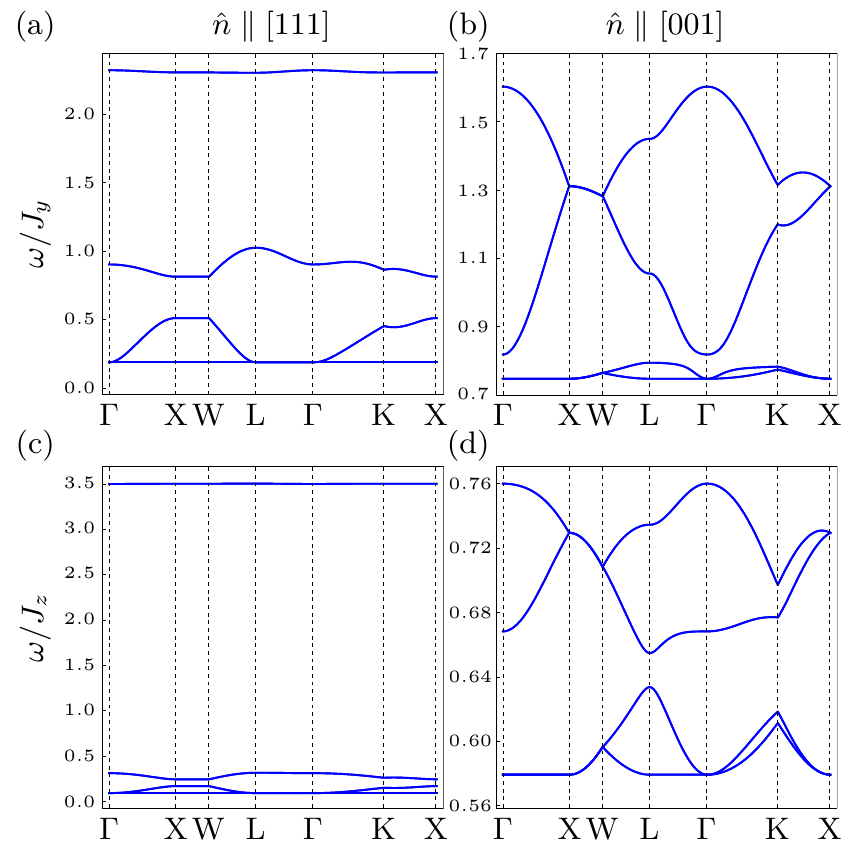}
    \caption{Linear spin wave spectra under external magnetic fields 
    along $[111]$ (a,c) and $[001]$ (b,d) direction. The corresponding 
    classical ground states are ``1-in 3-out'' and ``2-in 2-out'', respectively. 
    (a-b) In the octupolar U(1) QSL regime, we set $J_y$ as the energy unit 
    and fix ${J^y_{\pm}=-0.075J_y}$ and ${J^y_{\pm\pm}=0.025J_y}$. 
 (c-d) In the dipolar U(1) QSL regime, the energy unit is set to be ${J_z=1.0}$ 
    with ${J^z_{\pm}=-0.075J_z}$ and $J^y_{\pm\pm}=-0.025J_z$. 
    The strength of external field is fixed to ${h=4.0J_y}$ (${h=4.0J_z}$) 
    in the octupolar (dipolar) regime to ensure the correct spin configuration. 
    We set $\theta=0$.}
    \label{spin_wave}
\end{figure}

\begin{figure}[th]
    \centering
    \includegraphics[width=0.47\textwidth]{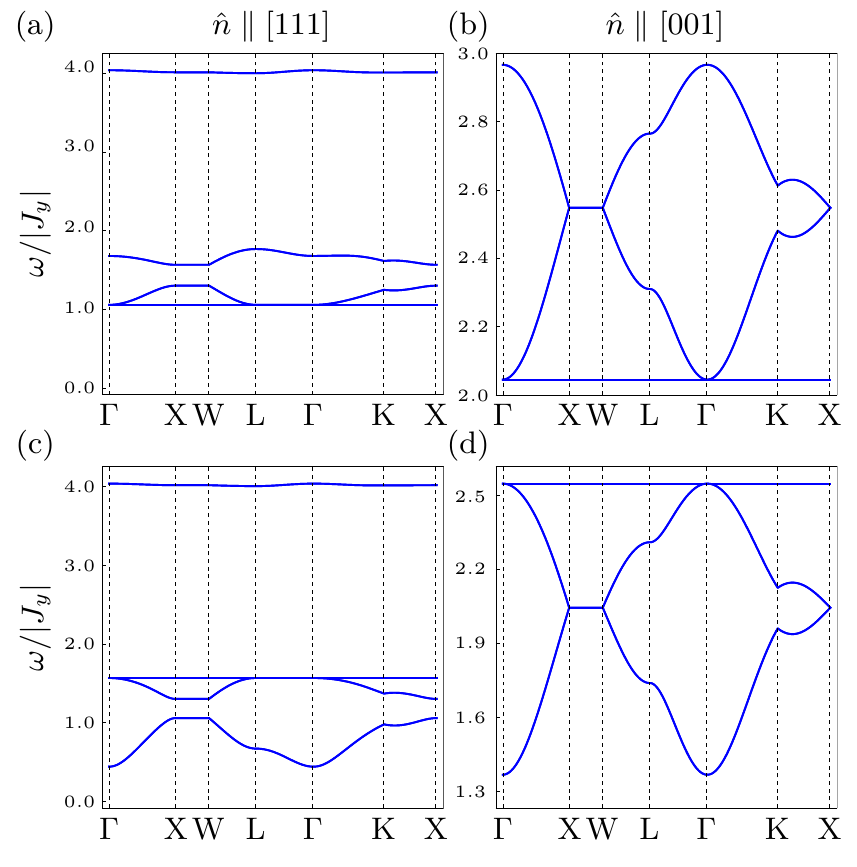}
    \caption{Linear spin wave spectra under external magnetic fields 
    along $[111]$ (a,c) and $[001]$ (b,d) directions. 
    we take ${J_x=J_z=J_{xz}=0}$ and the magnetic field strength 
    $h=8.0|J_y|$ to ensure the ``1-in 3-out'' and ``2-in 2-out'' 
    classical spin ground states, respectively. 
    In (a-b) we set ${J_y=1}$ as the energy unit and 
    in (c-d) we set ${J_y=-1}$.}
    \label{spin_wave_Jy}
\end{figure}

To illustrate the above thoughts, we proceed to calculate the spin
wave dispersions for the parameter choices of the dipolar U(1) QSL 
and the octupolar U(1) QSL, respectively. In practice, one obtains 
the spin wave spectra from the neutron scattering measurement by 
applying magnetic fields to polarize the spin and then extract  
the couplings. Since the experiments are not available yet, we
choose the representative parameters for the dipolar U(1) QSL 
and the octupolar U(1) QSL, and perform our spin wave analysis. 
To carry out the actual calculation, we invoke the well-known
Holstein-Primakoff spin wave theory to expand the spin operator.  
We first consider the application of the magnetic field along the [111]
direction. In the strong field limit, the spin configuration 
would simply be a ``3-in 1-out'' state. For our parameter choices 
that are given in Fig.~\ref{spin_wave} and Fig.~\ref{spin_wave_Jy},
it is legitimate to express the spin operators of the 0-th 
sublattice as, 
\begin{eqnarray}
    S_i^{+} & = & b_i^{},\quad
    S_i^{-} = b_i^{\dagger}, \\
    S_i^z & = & {1}/{2} - b_i^{\dagger} b_i^{},
\end{eqnarray}
and for the remaining three sublattices, we have 
\begin{eqnarray}
    S_i^{+} & = & b_i^{\dagger},\quad
    S_i^{-} = b^{}_i, \\
    S_i^z & = & -{1}/{2} + b_i^{\dagger} b_i^{}. 
\end{eqnarray}

After substituting $S_i$ in Eq.~\eqref{eq1} with the bosonic creation 
(annihilation) operators $b_i^{\dagger}$ ($b_i$) and then performing the
Fourier transformation
\begin{equation}
    b_i=\frac{1}{\sqrt{L}}\sum_{\bm{k}\in \text{B.Z.}} b_{s}(\bm{k})e^{i\bm{k}\cdot \bm{r}_l},
\end{equation}
where $\bm{r}_l$ is the position vector of unit cell containing magnetic ion 
$i$ and $s$ refers to the corresponding sublattice index, the XYZ model 
Hamiltonian under the magnetic field can be recast in terms of boson bilinears as 
\begin{equation}
    H_{sw}= \sum_{\bm{k}\in \text{B.Z.}} \bm{b}^{\dagger}(\bm{k})h(\bm{k})\bm{b}(\bm{k})
     + H_{\text{Zeeman}}(\bm{k}).
\end{equation}
Here, $\bm{b}(\bm{k})$ is a set of bosonic operator basis $\bm{b}(\bm{\bm{k}})=[b_0(\bm{k}),
\ldots, b_3(\bm{k}),b_0^{\dagger}(-\bm{k})\ldots,b_3^{\dagger}(-\bm{k})]^{T}$ and $h(\bm{k})$ 
is a $8 \times 8$ Hermitian matrix that can be written in the block form as
\begin{equation}
    h(\bm{k})=
    \begin{pmatrix}
        A(\bm{k}) & B(\bm{k}) \\
        B^{\dagger}(\bm{k}) & \bar{A}(-\bm{k})
    \end{pmatrix}
    +E_{\text{cl}},
\end{equation}
where $E_{\text{cl}}$ is the classical ground state energy. 
The matrix elements $A(\bm{k})$ and $B(\bm{k})$ are defined as
\begin{widetext}
\begin{equation}
    A(\bm{k})= 
    \begin{pmatrix}
        \frac{3}{2}J_z & \frac{J_x-J_y}{8}\big[e^{-\frac{i}{2}(k_y+k_z)}+1\big] & \frac{J_x-J_y}{8}\big[e^{-\frac{i}{2}(k_x+k_z)}+1\big] & \frac{J_x-J_y}{8}\big[e^{-\frac{i}{2}(k_x+k_y)}+1\big] \\
        \frac{J_x-J_y}{8}\big[e^{\frac{i}{2}(k_y+k_z)}+1\big] & -\frac{1}{2}J_z & \frac{J_x+J_y}{8}\big[e^{\frac{i}{2}(k_y-k_x)}+1\big] & \frac{J_x+J_y}{8}\big[e^{\frac{i}{2}(k_z-k_x)}+1\big] \\
        \frac{J_x-J_y}{8}\big[e^{\frac{i}{2}(k_x+k_z)}+1\big] & \frac{J_x+J_y}{8}\big[e^{\frac{i}{2}(k_x-k_y)}+1\big] & -\frac{1}{2}J_z & \frac{J_x+J_y}{8}\big[e^{\frac{i}{2}(k_z-k_y)}+1\big] \\
        \frac{J_x-J_y}{8}\big[e^{\frac{i}{2}(k_x+k_y)}+1\big] & \frac{J_x-J_y}{8}\big[e^{\frac{i}{2}(k_x-k_z)}+1\big] & \frac{J_x-J_y}{8}\big[e^{\frac{i}{2}(k_y-k_z)}+1\big] & -\frac{1}{2}J_z
    \end{pmatrix}
\end{equation}
and 
\begin{equation}
    B(\bm{k})=
    \begin{pmatrix}
        0 & \frac{J_x+J_y}{8}\big[e^{-\frac{i}{2}(k_y+k_z)}+1\big] & \frac{J_x+J_y}{8}\big[e^{-\frac{i}{2}(k_x+k_z)}+1\big] & \frac{J_x+J_y}{8}\big[e^{-\frac{i}{2}(k_x+k_y)}+1\big] \\
        \frac{J_x+J_y}{8}\big[e^{\frac{i}{2}(k_y+k_z)}+1\big] & 0 & \frac{J_x-J_y}{8}\big[e^{\frac{i}{2}(k_y-k_x)}+1\big] & \frac{J_x-J_y}{8}\big[e^{\frac{i}{2}(k_z-k_x)}+1\big] \\
        \frac{J_x+J_y}{8}\big[e^{\frac{i}{2}(k_x+k_z)}+1\big] & \frac{J_x-J_y}{8}\big[e^{\frac{i}{2}(k_x-k_y)}+1\big] & 0 & \frac{J_x-J_y}{8}\big[e^{\frac{i}{2}(k_z-k_y)}+1\big] \\
        \frac{J_x+J_y}{8}\big[e^{\frac{i}{2}(k_x+k_y)}+1\big] & \frac{J_x+J_y}{8}\big[e^{\frac{i}{2}(k_x-k_z)}+1\big] & \frac{J_x+J_y}{8}\big[e^{\frac{i}{2}(k_y-k_z)}+1\big] & 0
    \end{pmatrix}.
\end{equation}
\end{widetext}
$\bar{A}(\bm{k})$ is the complex conjugate of $A(\bm{k})$.

The Zeeman coupling in this approximation becomes
\begin{multline}
    H_{\text{Zeeman}}(\bm{k}) = -h \sum_{\bm{k} \in \text{B.Z.}} \sum_{s=0}^{3} \hat{n} \cdot \hat{z}_s \\
    \times [b_s^{\dagger}(\bm{k}) b_s(\bm{k}) + b_s(-\bm{k}) b_s^{\dagger}(-\bm{k})].
\end{multline}

In our illustrative calculations, we keep ${\theta=0}$ as the previous sections for simplicity. 
Without losing generality, we set ${J_y=1.0}$ and ${J_z=1.0}$ as the energy unit for 
the octupolar and dipolar U(1) QSLs, respectively. Other parameters are set to be 
${J^y_{\pm}=-0.075J_y}$, 
${J^y_{\pm\pm}=0.025J_y}$ and ${J^z_{\pm}=-0.05J_z}$, ${J^z_{\pm\pm}=-0.025J_z}$ 
in order to guarantee the predominance of $\tilde{J}_y$ or $\tilde{J}_z$ in each case. 
In both cases, the strength of the magnetic field is fixed to ${h=4.0 J_y}$ (or ${h=4.0 J_z}$) 
which is strong enough to ensure the magnetic ground state is a ``1-in 3-out'' 
spin configuration. Diagonalizing the quadratic Hamiltonian $H_{sw}$, one can 
obtain the linear spin wave spectrum as depicted in Fig.~\ref{spin_wave}(a) and (c). 
In both the octupolar (a) and dipolar (c) regime, the degenerate bands with the highest 
energies originate from the deviation of the spin whose local $z$ component is 
paralleled to the field direction. In the dipolar U(1) QSL, 
this band is nearly flat and there is a huge gap between other bands 
while the energy gap is moderate in the octupolar one.

With the same parameters but an external magnetic field along $[001]$ direction, 
a ``2-in 2-out'' spin configuration is favored. In this ground state, 
the spin operators of the 0-th and 3-rd sublattice can be expressed as, 
\begin{eqnarray}
    S_i^{+} & = & b_i^{},\quad
    S_i^{-} = b_i^{\dagger}, \\
    S_i^z & = & {1}/{2} - b_i^{\dagger} b_i^{},
\end{eqnarray}
and for the remaining two sublattices, we have
\begin{eqnarray}
    S_i^{+} & = & b_i^{\dagger},\quad
    S_i^{-} = b^{}_i, \\
    S_i^z & = & -{1}/{2} + b_i^{\dagger} b_i^{}. 
\end{eqnarray}
As shown in Fig.~\ref{spin_wave}, the bandwidth of linear spin wave spectrum 
(Fig.~\ref{spin_wave}(b) and (d)) is significantly smaller than the previous case. 
This is because none of the four local $\hat{z}$ directions in one magnetic unit cell 
is parallel to the external field. There is a reduced energy cost of spin flipping. 
In the dipolar regime as shown in Fig.~\ref{spin_wave}(d), the bandwidth is much 
smaller by comparison.

In order to further study the difference between the dipolar and octupolar U(1) QSLs 
from the perspective of the spin wave excitations, we take the special case with 
${J_x=J_z=J_{xz}=0}$ and calculate the spin wave spectrum of a $J_y$ related Hamiltonian 
\begin{equation}
    H=\sum_{\langle ij \rangle}J_y S_i^y S_j^y-h\sum_{i}(\hat{n}\cdot \hat{z}_i)S_i^z.
\end{equation}
Here, we set the energy unit to be ${|J_y|=1}$ but the sign of $J_y$ can be changed. 
To ensure the same classical spin ground states as we have discussed above, the strength 
of external magnetic field is fixed to be ${h=8|J_y|}$. We show the spin wave results 
in Fig.~\ref{spin_wave_Jy} where ${J_y=1}$ for (a-b) and ${J_y=-1}$ for (c-d).
Our results could potentially provide a guidance for the future inelastic neutron 
scattering measurements in the strong magnetic fields.

\section{Discussion}
\label{sec6}

The Ce-based pyrochlore QSL materials 
(Ce$_2$Sn$_2$O$_7$ and Ce$_2$Zr$_2$O$_7$) represent a family of QSL 
materials whose models are provided theoretically~\cite{huang1311quantum,li1607symmetry}. 
The major task would be to establish connections between the theoretical 
results/understanding and experiments. The main result in this paper
is based on the U(1) QSLs with dipole-octupole doublets, 
and the experimental predictions are the spectroscopic 
properties. It has been shown that the spinon spectrum 
could have an enhanced spectral periodicity with a folded Brillouin zone
and the proximate orders could break the lattice translational symmetry by 
doubling the unit cell. Another set of experiments would be thermal Hall transports. 
As we will explain in a separate paper~\cite{xiaotian} that focuses on thermal 
Hall effect, we predict that there should be a non-trivial topological thermal 
Hall effect for ``magnetic monopoles'' due to the dual Berry phase effect 
in the dipolar U(1) QSL (or any other spin-ice based U(1) QSL materials) 
while there is no such topological thermal Hall effect for the 
``magnetic monopoles'' excitations in the octupolar U(1) QSL. 
The possibility of $\mathbb{Z}_2$ QSL is not considered here. 
Although the region of possible $\mathbb{Z}_2$ QSL is tiny on 
the unfrustrated (sign-problem-free) side~\cite{huang1806extended}, 
the presence of $\mathbb{Z}_2$ QSL on the frustrated side is not so 
clear. Thus, $\mathbb{Z}_2$ QSL may still be possible, and the 
spectrum would be fully gapped. This may be examined carefully 
with the detailed specific heat measurements.

For the XYZ model on the pyrochlore lattice, it is ready to see that the model 
reduces to the Heisenberg model when all three couplings are equal. The ground 
state of the pyrochlore lattice Heisenberg model is one of the hardest problems 
in quantum magnetism. From the property of the XYZ model, one could at least 
conclude that the ground state for the Heisenberg model cannot be the
$\pi$-flux U(1) QSL for the XXZ model in the frustrated regime. 
This is because the three spin components have different 
physical meanings in the emergent spinon-gauge description while
the three spin components are symmetrically related by 
the SU(2) spin rotation at the Heisenberg point.


\section*{Acknowledgments}

We acknowledge an anonymous refereee for his/her suggestion of 
a third possibility of field-driven transition. 
We acknowledge Mike Hermele from University of Colorado Boulder 
and Yi-Ping Huang from MPI-Dresden for a previous collaboration. 
We acknowledge Chenjie Wang for discussion. This work is supported 
by the Ministry of Science and Technology of China with Grant 
No.2016YFA0301001, 2016YFA0300501, 2018YFE0103200 and by the 
General Research Fund (GRF) No.17303819 from the Research Grant 
Council of Hong Kong.

\appendix

\section{Coordinate system\label{appA}}

\begin{figure}[b]
	\includegraphics[width=5.5cm]{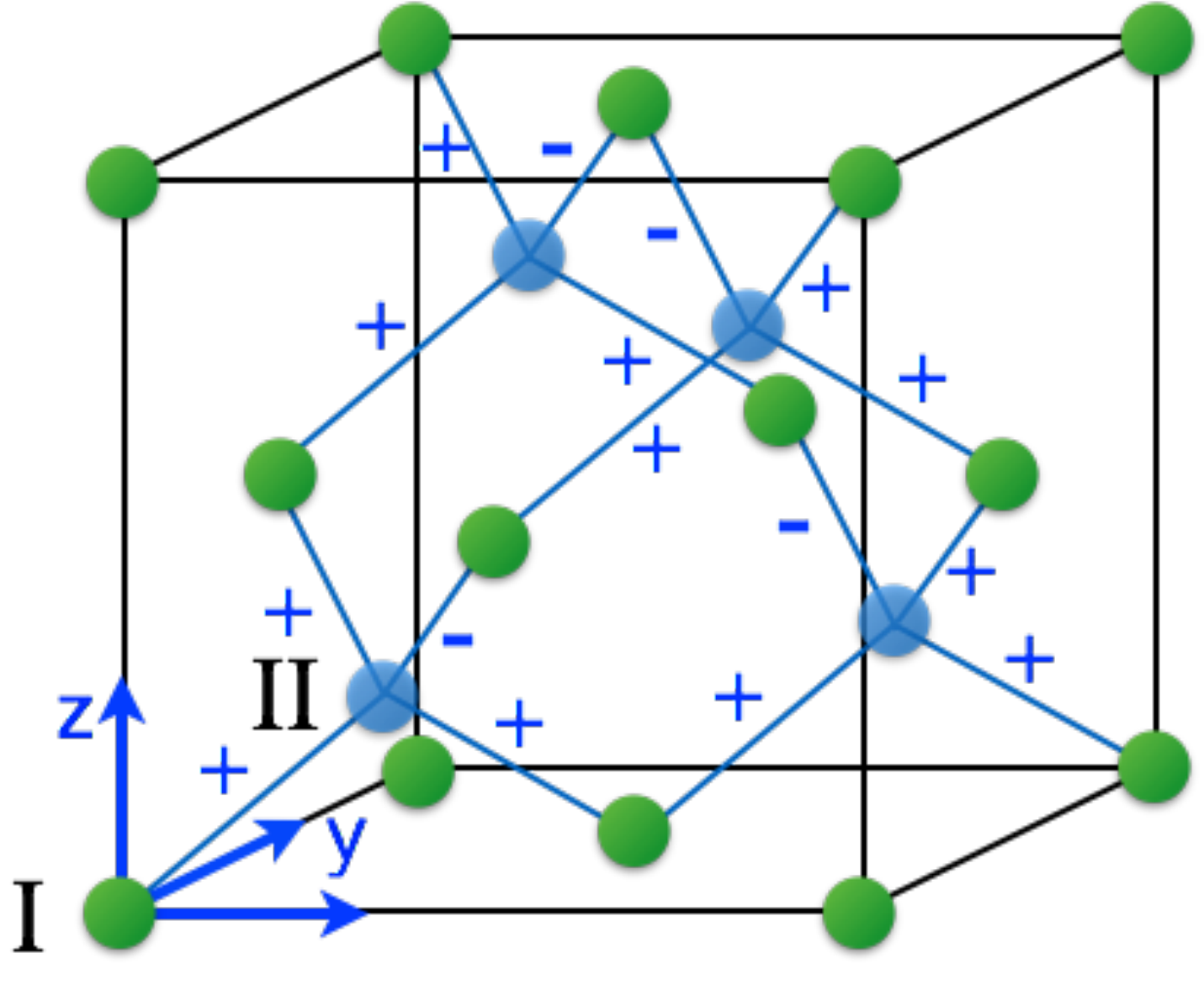}
\caption{(Color online.)
The diamond lattice formed by the tetrahedral centers of the pyrochlore lattice. 
The choice of spinon hopping is labelled in the plot.
}
    \label{fig:lattice}
\end{figure}

The centers of the corner-sharing tetrahedra in the pyrochlore lattice 
constitute a diamond structure with two sublattices, which we denote I and II; 
see Fig.~\ref{fig:lattice}. We choose the origins of the two sublattices as follows,
\envelope{eqnarray}{
    \bs{O}_{\rm I} &=& \(0, 0, 0\), \\
    \bs{O}_{\rm II} &=& \(\frac{1}{4}, \frac{1}{4}, \frac{1}{4}\).
}
The basis vectors of the diamond lattices are taken to be
\envelope{eqnarray}{
    \bs{a}_1 &=& \(0, \frac{1}{2}, \frac{1}{2}\), \\
    \bs{a}_2 &=& \(\frac{1}{2}, 0, \frac{1}{2}\), \\
    \bs{a}_3 &=& \(\frac{1}{2}, \frac{1}{2}, 0\).
}

\begin{table}[t]
\begin{tabular}{p{1cm} p{1.5cm} p{1.5cm} p{1.5cm} p{1.5cm}}
\hline\hline
    $\mu$ &  0 & 1 & 2 & 3 \\
    $\hat{x}_\mu$ & $\frac{1}{\sqrt{2}}[\bar{1} 1 0]$ &  $\frac{1}{\sqrt{2}}[\bar{1} \bar{1} 0]$ & $\frac{1}{\sqrt{2}}[1 1 0]$ & $\frac{1}{\sqrt{2}}[1 \bar{1} 0]$\\
    $\hat{y}_\mu$ & $\frac{1}{\sqrt{6}}[\bar{1} \bar{1} 2]$ &  $\frac{1}{\sqrt{6}}[\bar{1} 1 \bar{2}]$ & $\frac{1}{\sqrt{6}}[1 \bar{1} \bar{2}]$ & $\frac{1}{\sqrt{6}}[1 1 2]$\\
    $\hat{z}_\mu$ & $\frac{1}{\sqrt{3}}[111]$ &  $\frac{1}{\sqrt{3}}[1\bar{1} \bar{1}]$ & $\frac{1}{\sqrt{3}}[\bar{1} 1 \bar{1}]$ & $\frac{1}{\sqrt{3}}[\bar{1} \bar{1} 1]$\\
\hline\hline
\end{tabular}
\caption{Local coordinate frames for the four sublattices on the pyrochlore lattice.
The local $\hat{x}$ and $\hat{y}$ directions are not really used in this work because the 
transverse spin components of local moments are octupolar moments and 
are defined in the internal spin space. }
\label{tab1}
\end{table}

For each site of the I (II) sublattice there are four vertices of the II (I) sublattice that are nearest neighbors to it, with displacement vectors 
\envelope{eqnarray}{
    \eta_{\bs{r}} \, \bs{e}_0 &=& \frac{\eta_{\bs{r}}}{4} \(+1, +1, +1\), \\
    \eta_{\bs{r}} \, \bs{e}_1 &=& \frac{\eta_{\bs{r}}}{4} \(+1, -1, -1\), \\
    \eta_{\bs{r}} \, \bs{e}_2 &=& \frac{\eta_{\bs{r}}}{4} \(-1, +1, -1\), \\
    \eta_{\bs{r}} \, \bs{e}_3 &=& \frac{\eta_{\bs{r}}}{4} \(-1, -1, +1\),
}
where $\eta_{\bs{r}} = \pm 1$ for $\bs{r}$ in sublattice I and II, respectively.
At the midpoint of each of such bonds, there is a vertex of the pyrochlore lattice.
Correspondingly, we define the local coordinate systems on the four sublattices of 
the pyrochlore lattice, as summarized in the Table~\ref{tab1}.

\vspace{1cm}
\section{Gauge pattern for octupolar U(1)$_\pi$ QSL state and Bloch hamiltonian \label{appB}}

As pointed out in the literature~\cite{lee1204generic, chen1706dirac}, 
in the frustrated regime $J_\pm < 0$ of the XYZ model 
(see Eqs.~(\ref{H_z}, \ref{H_y})) the ground state has $\pi$-flux within an elementary hexagon.
Within the gauge mean field theory, recall that the $s^{\pm}$ operators are gauge string operators, $s^{\pm}_{\bs{r}, \bs{r}^\p} = \frac{1}{2} e^{\pm i A_{\bs{r}, \bs{r}^\p} }$. 
We take the following gauge choice for the $\pi$-flux state,
\envelope{eqnarray}{
    A_{\bs{r}, \bs{r} + \bs{e}_\mu} = \epsilon_\mu \bs{Q} \cdot \bs{r},
}
where $\epsilon = (0110)$, $\bs{Q} = 2\pi(100)$, and $\bs{r}$ belongs to the I sublattice, 
as illustrated in Fig.~\ref{fig:lattice}.

\begin{figure*}[t]
	\includegraphics[width=.3\textwidth]{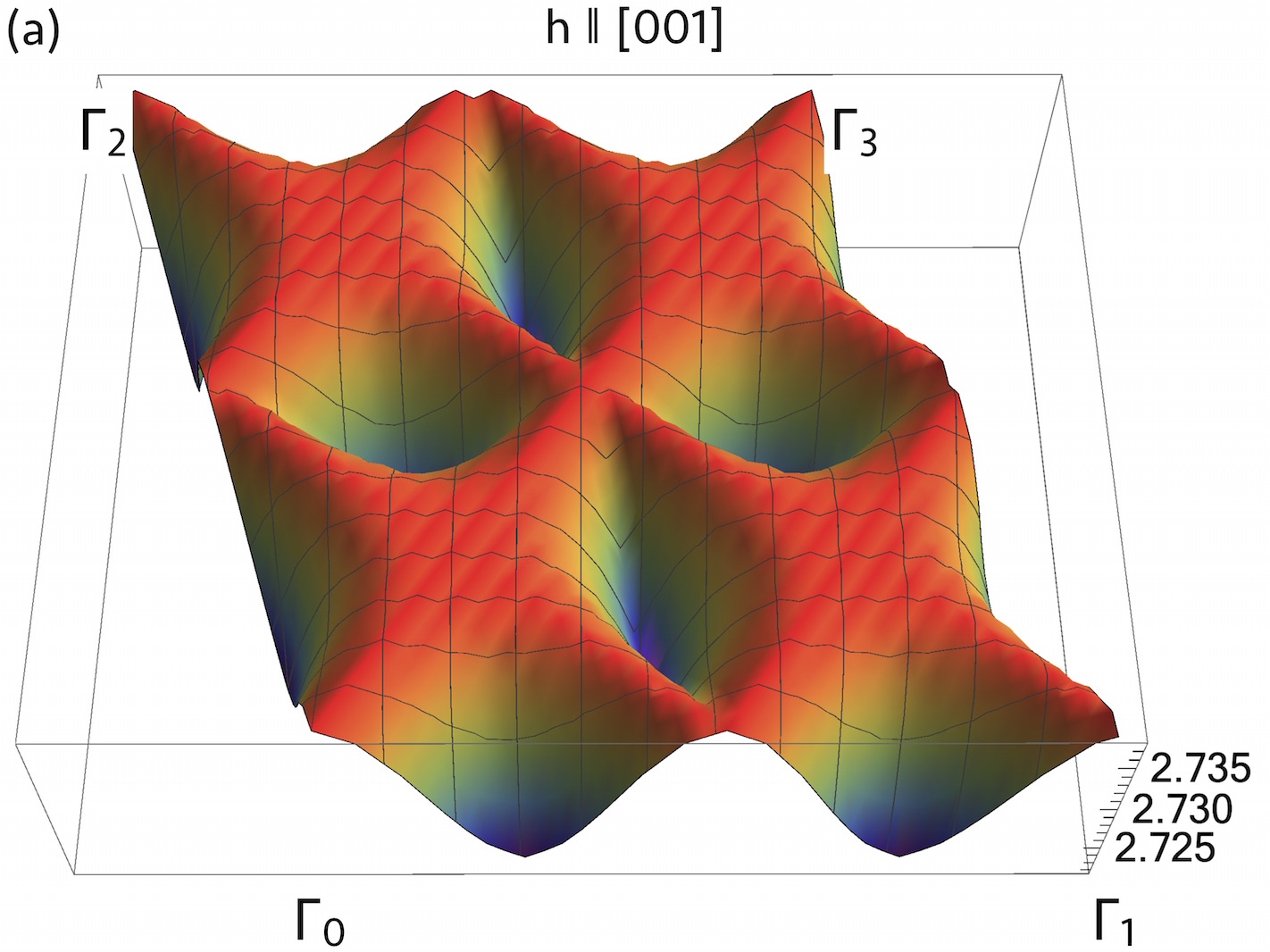}
    \includegraphics[width=.3\textwidth]{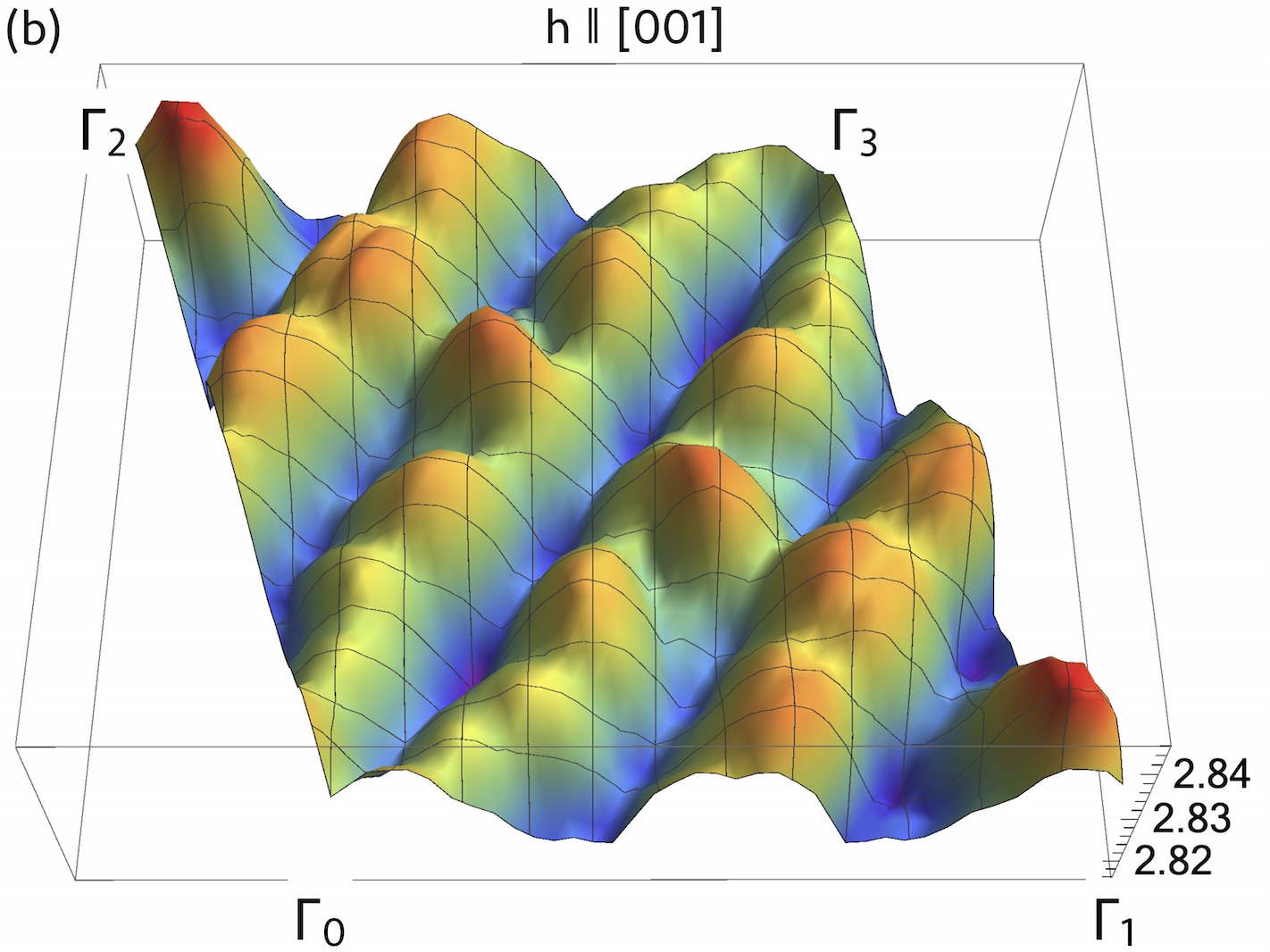}
    \includegraphics[width=.3\textwidth]{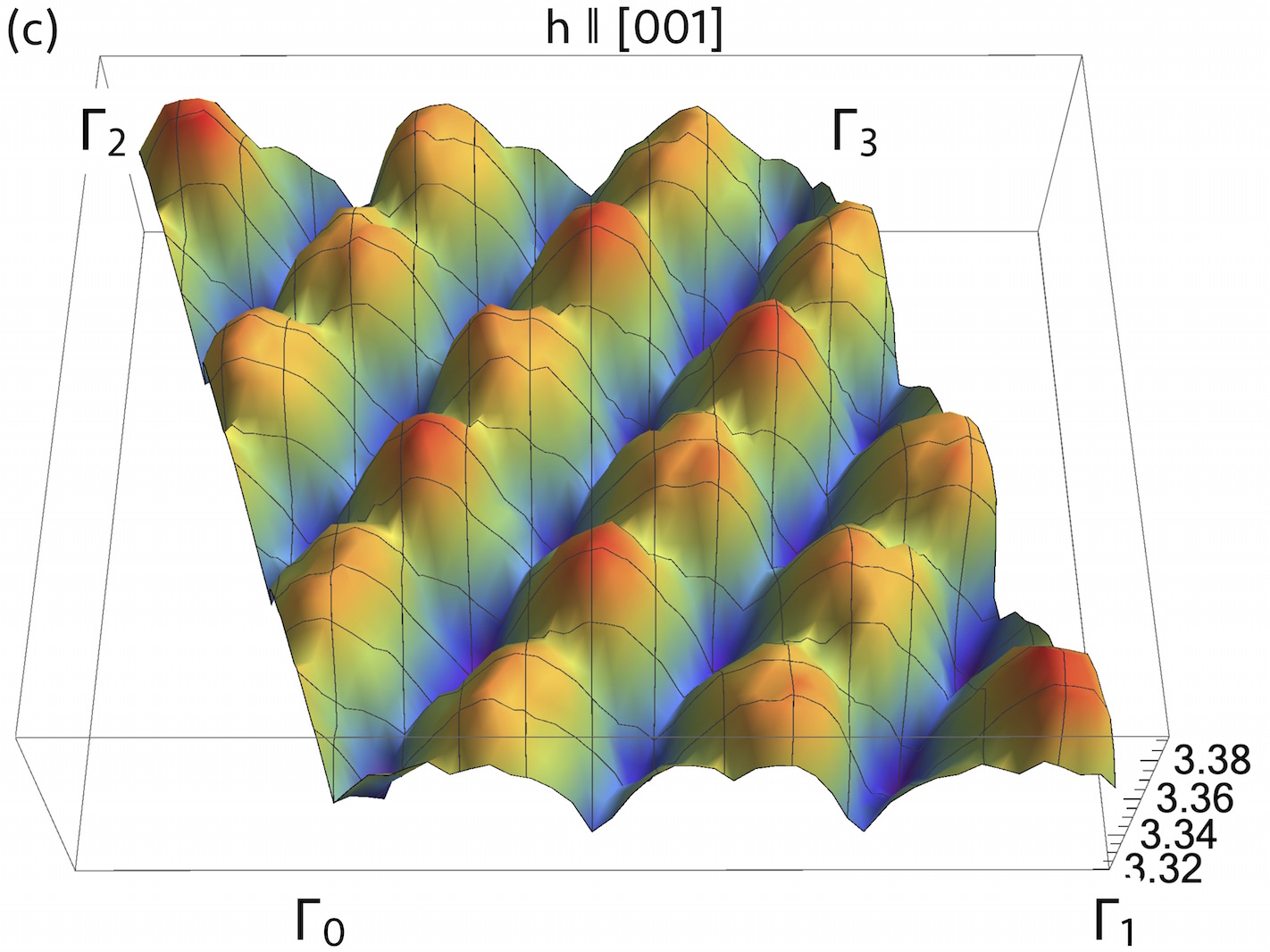}
	\includegraphics[width=.3\textwidth]{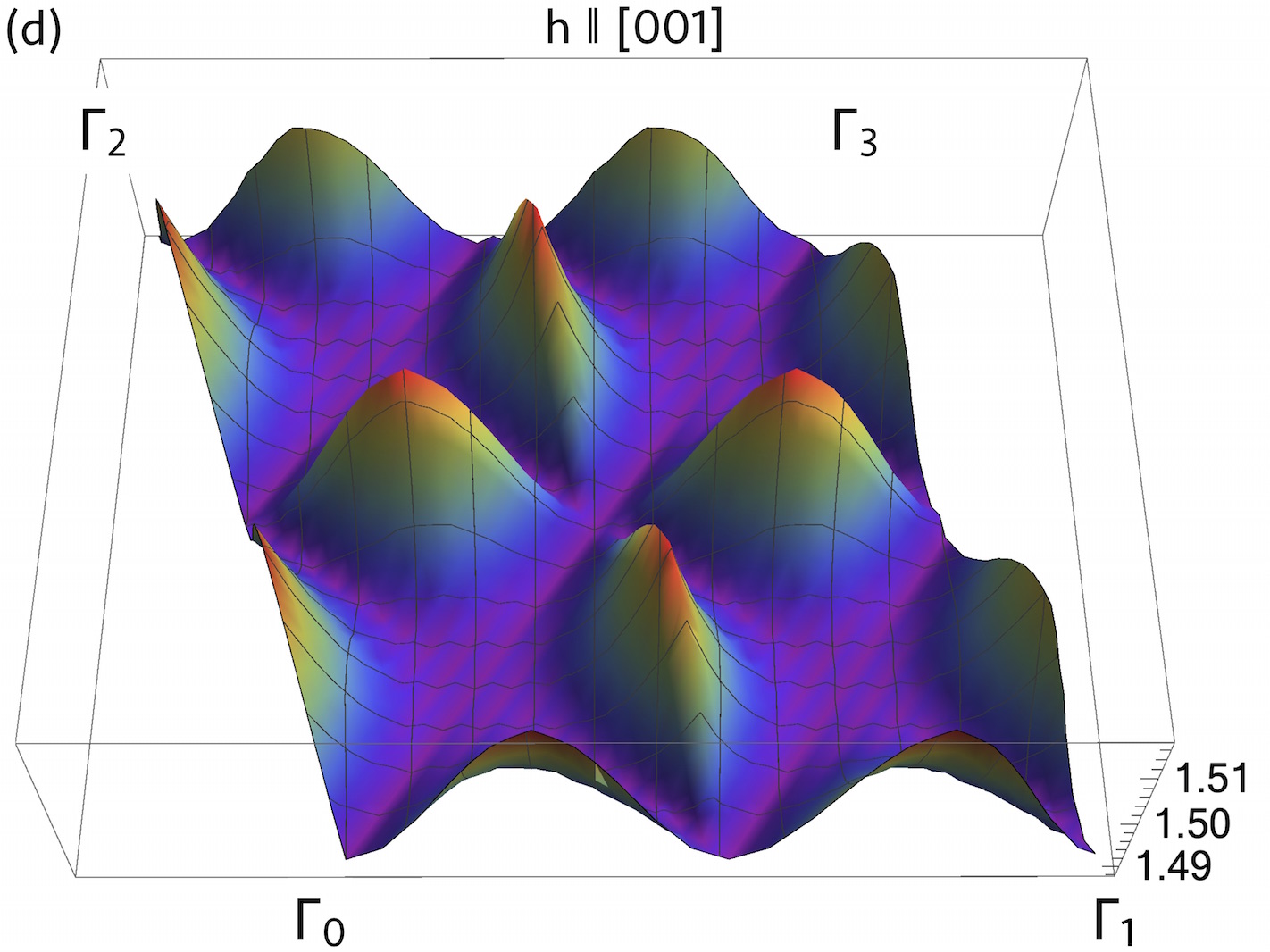}
    \includegraphics[width=.3\textwidth]{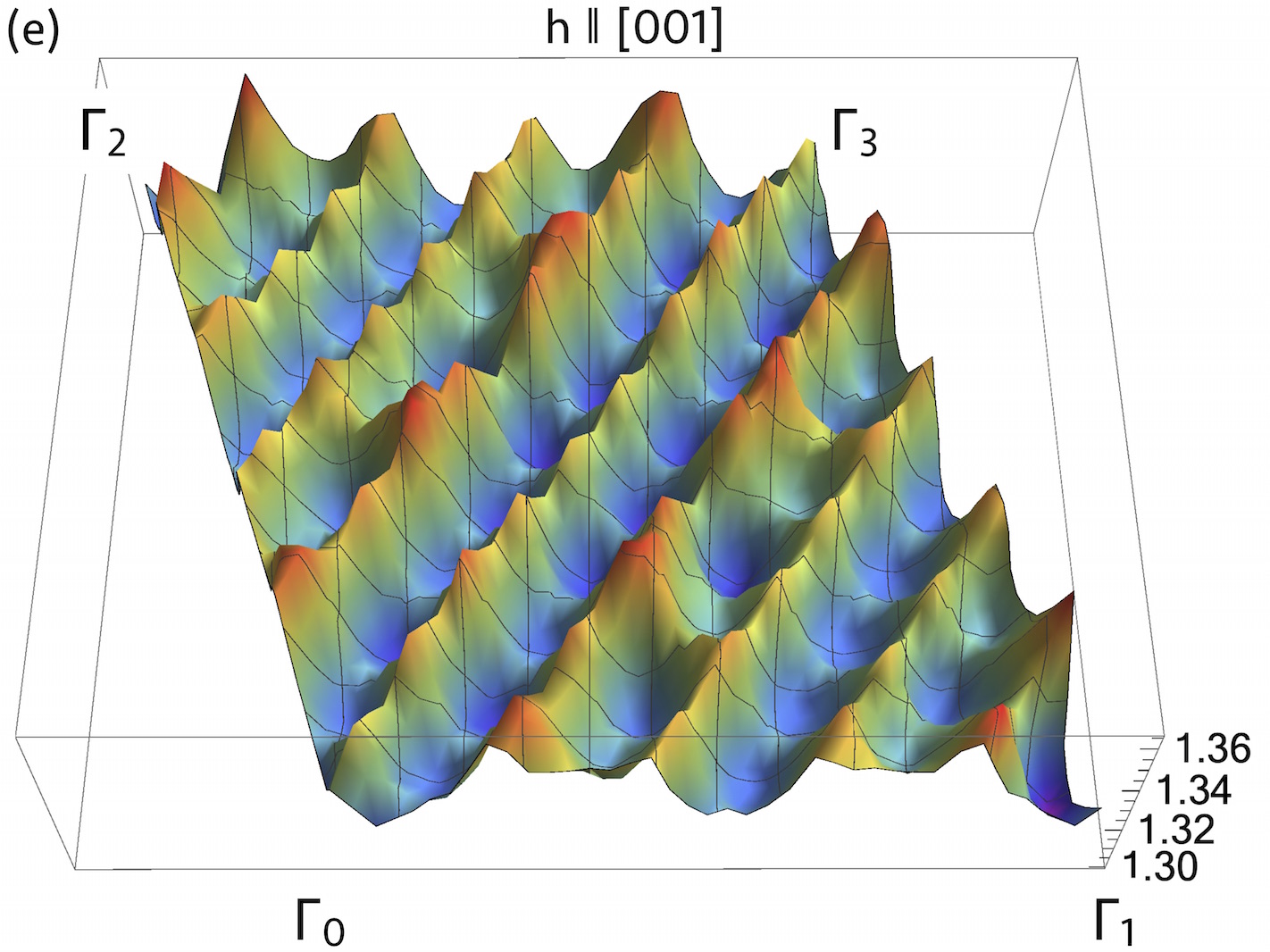}
    \includegraphics[width=.3\textwidth]{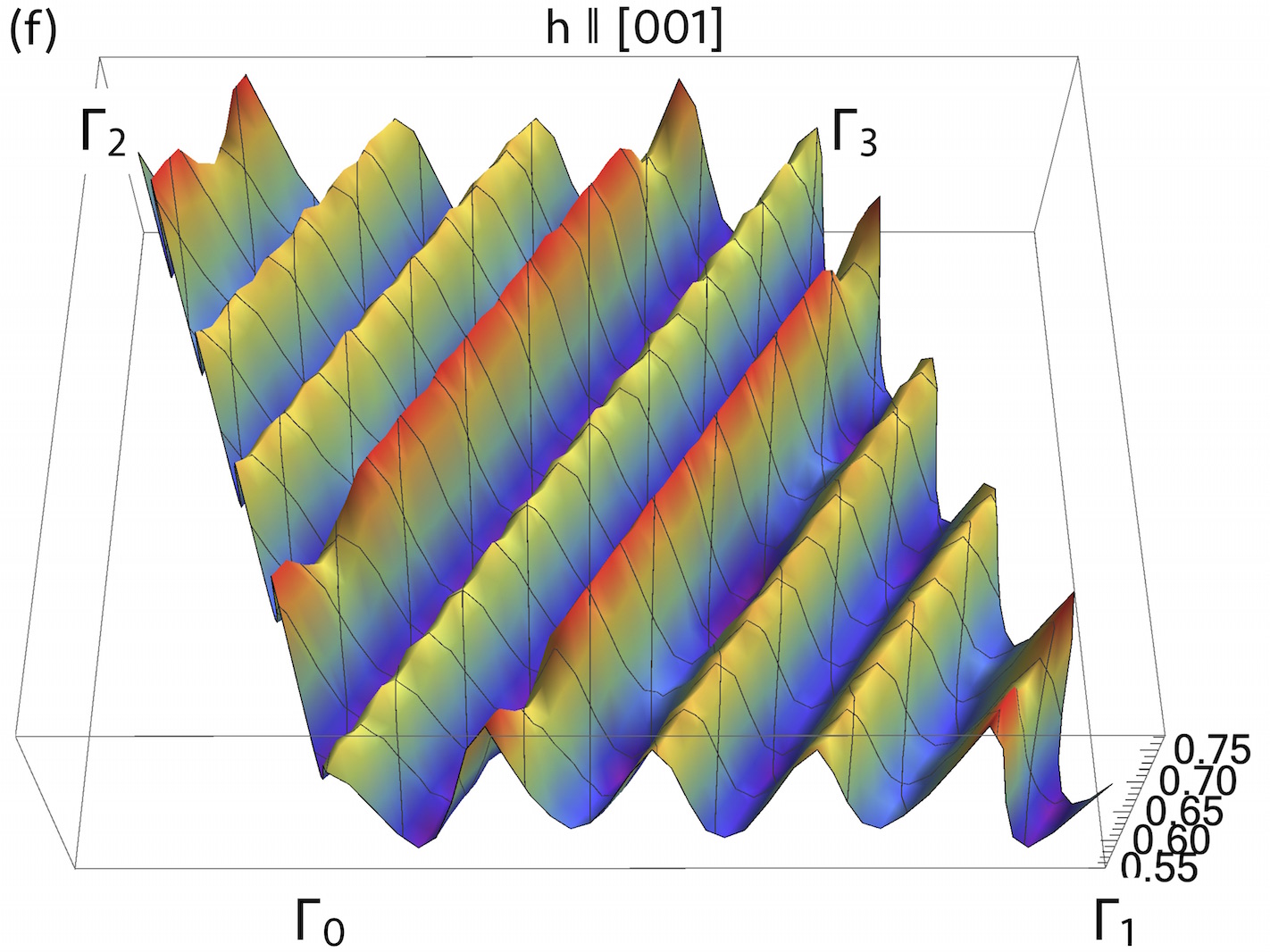}
\caption{(Color online.)
The upper (a, b, c) and lower (d, e, f) excitation edges of the two-spinon 
continuum in the octupolar U(1)$_\pi$ QSL, for fields along the [001] direction.
We fix $J_{\pm}^y = -0.33 J_y,J_{\pm\pm}^y =0, \theta = 0$, and take 
$h = 0.0 J_y$ in (a, d), $h = 0.2 J_y$ in (b, e), and $h = 1.0 J_y$ in (c, f).
}
    \label{fig:excitation_edge_001}
\end{figure*}


\begin{figure*}[htp]
	\includegraphics[width=.3\textwidth]{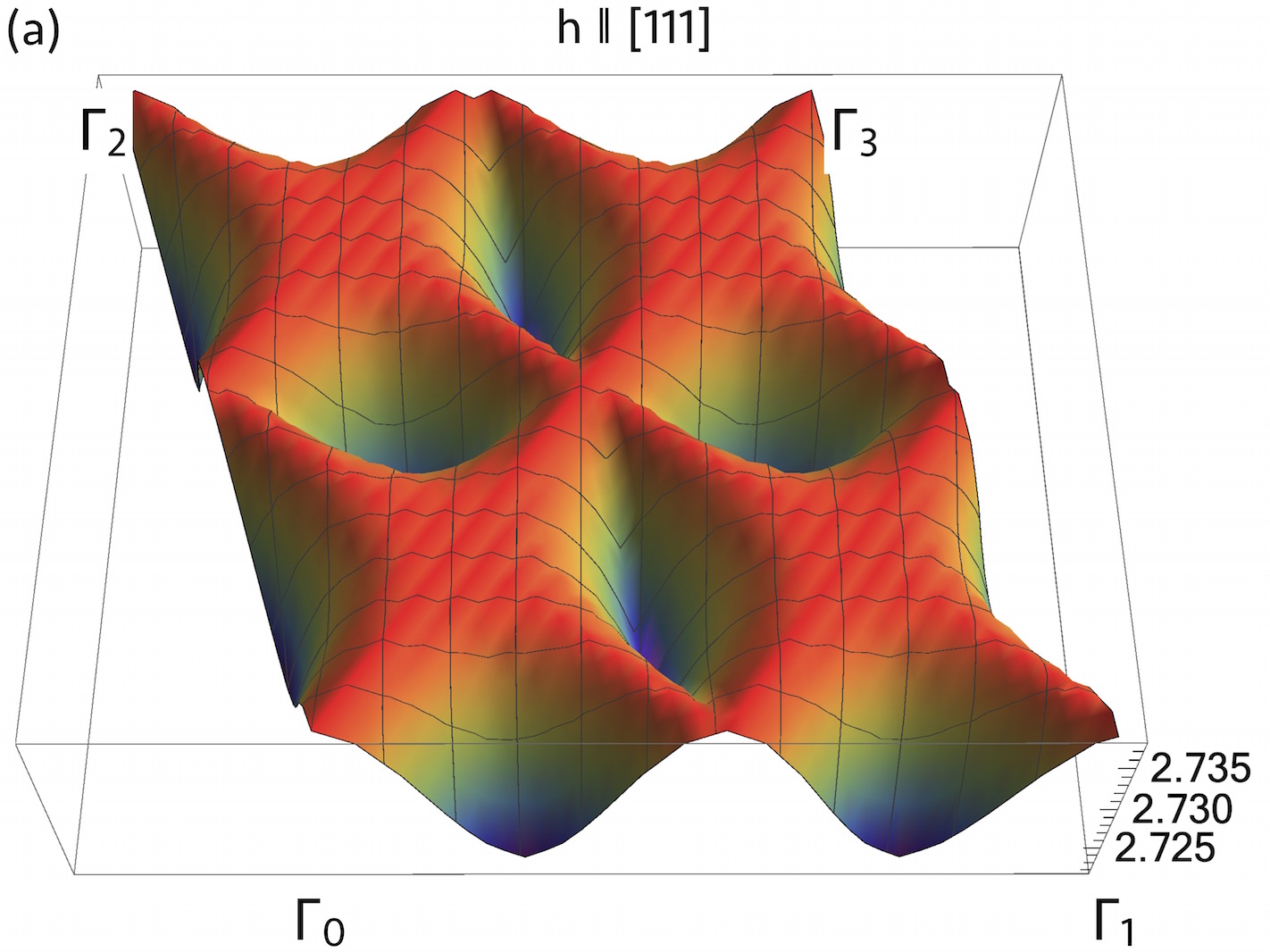}
    \includegraphics[width=.3\textwidth]{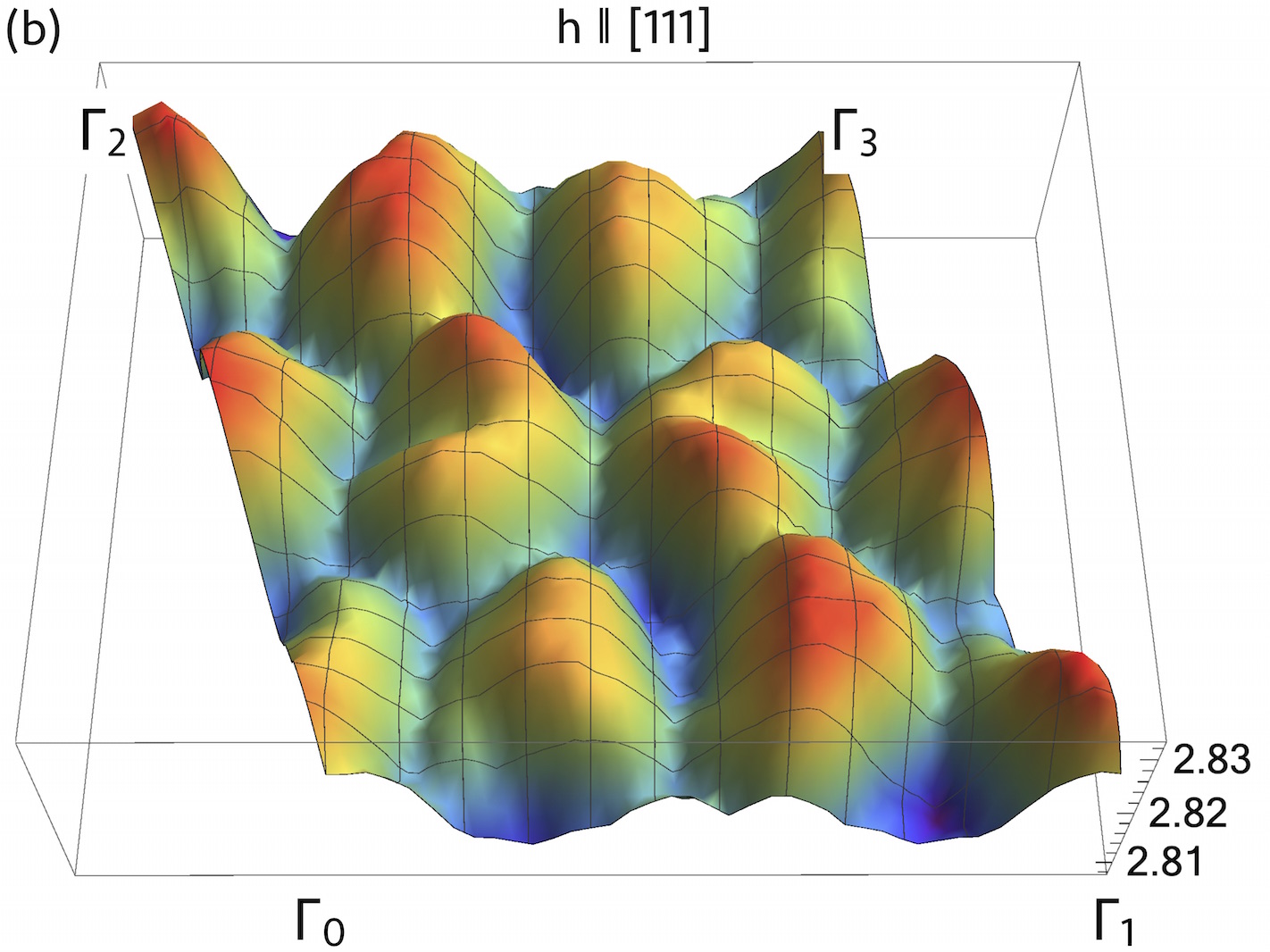}
    \includegraphics[width=.3\textwidth]{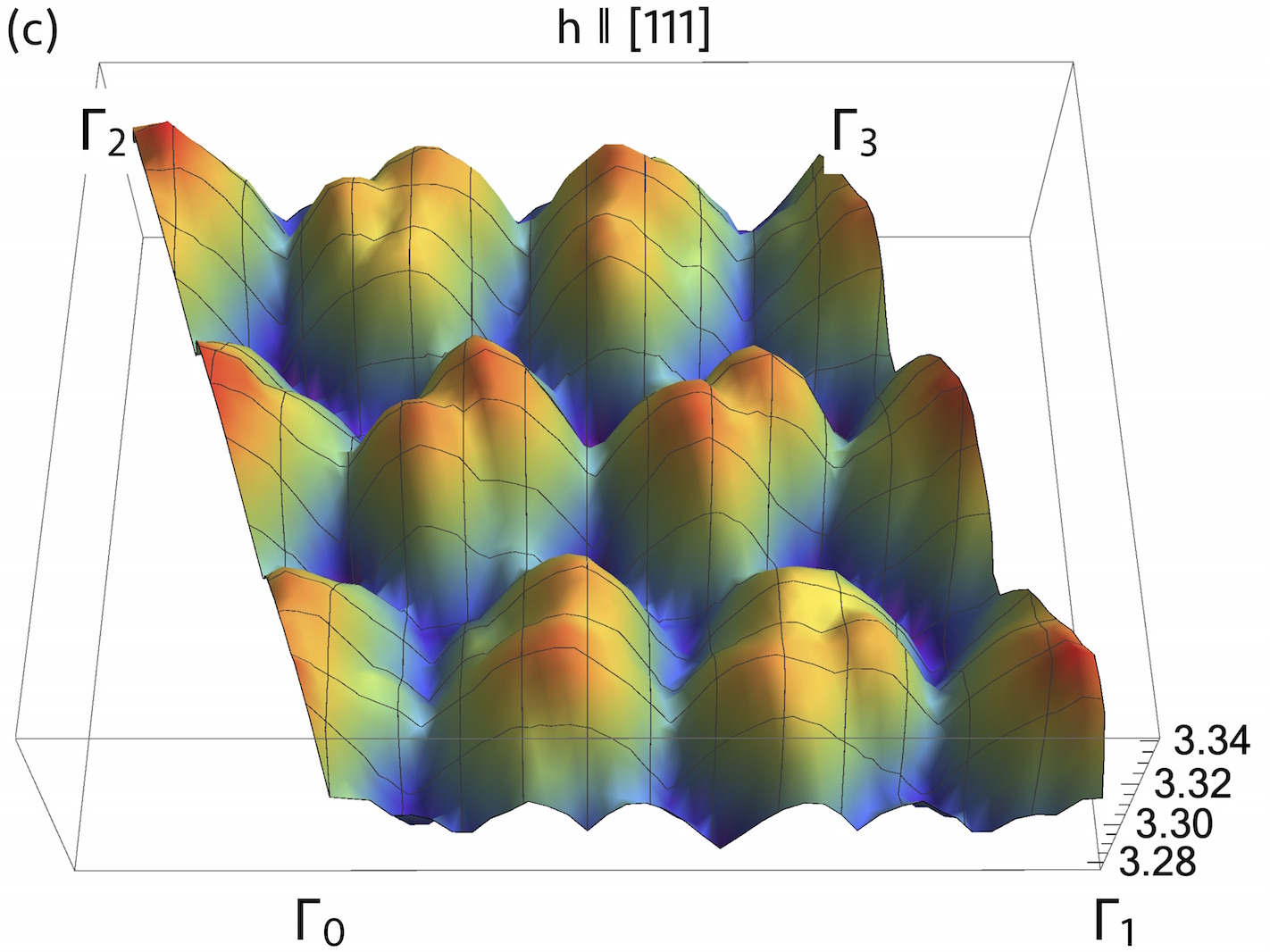}
	\includegraphics[width=.3\textwidth]{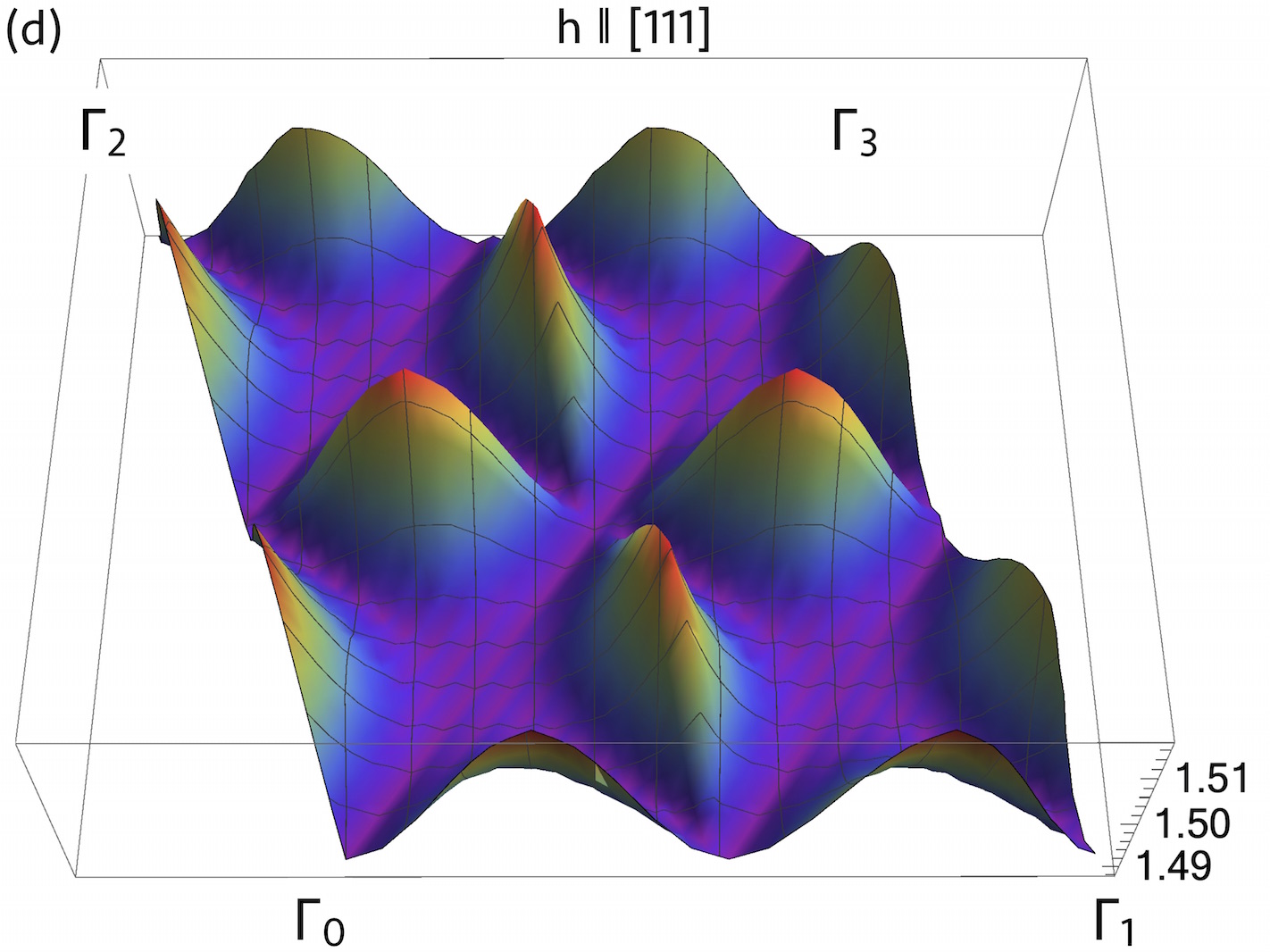}
    \includegraphics[width=.3\textwidth]{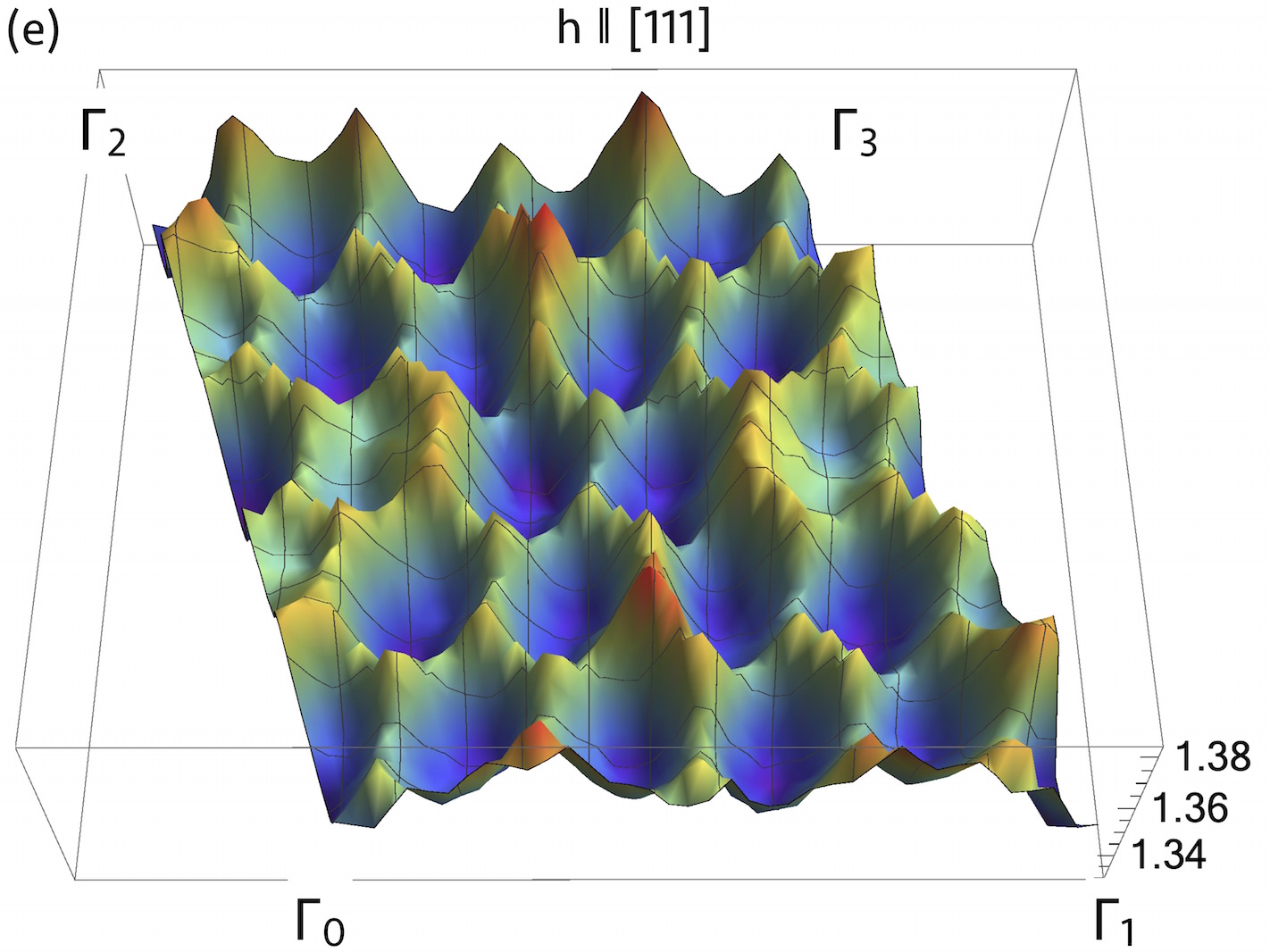}
    \includegraphics[width=.3\textwidth]{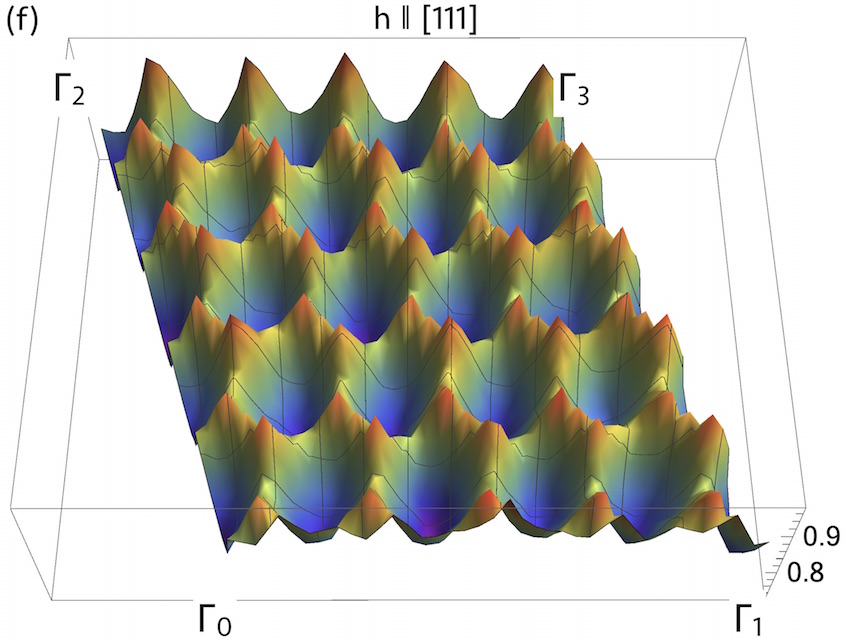}
\caption{(Color online.)
The upper (a, b, c) and lower (d, e, f) excitation edges of the two-spinon 
continuum in the octupolar U(1)$_\pi$ QSL, for fields along the [111] direction.
We fix $J_{\pm}^y = -0.33 J_y,J_{\pm\pm}^y =0, \theta =0 $, 
and take $h = 0.0 J_y$ in (a, d), $h = 0.2 J_y$ in (b, e), and $h = 1.0 J_y$ in (c, f).
}
    \label{fig:excitation_edge_111}
\end{figure*}

Because of the $\pi$-flux and the choice of $\bs{Q}$, the unit cell doubles in the $x$-direction.
Correspondingly, there are four sublattices of the system, which we term ${\rm I}, \pm$ and ${\rm II}, \pm$.
Specifically, a site on the origional I sublattice at $\bs{r}$ belongs to the ${\rm I}, +$ (${\rm I}, -$) sublattice if $\bs{Q} \cdot (\bs{r} - \bs{O}_{\rm I})$ is an even (odd) multiple of $\pi$; and this is similarly for a site in the II sublattice.

We focus on $H_y$ within the frustrated regime $J_\pm < 0$ and $\theta = 0$.
Under this fixed gauge, the spinon action is~\cite{lee1204generic}
\envelope{widetext}{
\envelope{eqnarray}{
	\mc{S} &=& \int d\tau \lz \sum_{\bs{r}} \frac{1}{2 \tilde{J}_y} \pd_\tau \Phi^\ast_{\bs{r}} \pd_\tau \Phi_{\bs{r}} + \lambda \sum_{\bs{r}} \(|\Phi_{\bs{r}}|^2 - 1\) 
	- J_{\pm} \sum_\bs{r} \sum_{\mu \neq \nu} \(
            \Phi^\dg_{\bs{r} + \eta_\bs{r} \bs{e}_\mu}
            \Phi^\pg_{\bs{r} + \eta_\bs{r} \bs{e}_\nu}
            s^{-\eta_\bs{r}}_{\bs{r}, \bs{r} + \eta_\bs{r} \bs{e}_\mu}
            s^{+\eta_\bs{r}}_{\bs{r}, \bs{r} + \eta_\bs{r} \bs{e}_\nu}
    \) \rz \nn
    &=& \sum_{\bs{k}} \int \frac{d\omega}{2\pi}\,
    	\boldsymbol{\Phi}^\dg_{\bs{k}}
		G_{\bs{k}, \omega}
		\boldsymbol{\Phi}^\pg_{\bs{k}}
}
where
\envelope{eqnarray}{
	G_{\bs{k}, \omega} = 
	\lz
		\( \frac{\omega^2}{2 \tilde{J}_y} + \lambda \) \bs{I}_{4\times 4} + 
		\envelope{pmatrix}{
			H_{\rm I, I}(\bs{k}) & H_{\rm I, II}(\bs{k}) \\
			H_{\rm II, I}(\bs{k}) & H_{\rm II, II}(\bs{k})
		}
	\rz, \quad
	\boldsymbol{\Phi}_{\bs{k}} = \envelope{pmatrix}{
			\Phi^\pg_{{\rm I},+}(\bs{k}) \\
			\Phi^\pg_{{\rm I},-}(\bs{k}) \\
			\Phi^\pg_{{\rm II},+}(\bs{k}) \\
			\Phi^\pg_{{\rm II},-}(\bs{k}) \\
		},
}
and
\envelope{eqnarray}{
	H_{\rm I, I}(\bs{k}) &=& - J_{\pm}
	\envelope{pmatrix}{
		+ \cos \frac{k_y}{2} \cos \frac{k_z}{2} & - \sin \frac{k_x}{2} \sin \frac{k_y}{2} - i \cos \frac{k_x}{2} \cos \frac{k_z}{2} \\
		- \sin \frac{k_x}{2} \sin \frac{k_y}{2} + i \cos \frac{k_x}{2} \cos \frac{k_z}{2} & -\cos \frac{k_y}{2} \cos \frac{k_z}{2}
	}, \\
	H_{\rm II, II}(\bs{k}) &=& - J_{\pm}
	\envelope{pmatrix}{
		-\sin \frac{k_y}{2} \sin \frac{k_z}{2} & + \cos \frac{k_x}{2} \cos \frac{k_y}{2} - i \cos \frac{k_z}{2} \sin \frac{k_x}{2} \\
		+ \cos \frac{k_x}{2} \cos \frac{k_y}{2} + i \cos \frac{k_z}{2} \sin \frac{k_x}{2} & +\sin \frac{k_y}{2} \sin \frac{k_z}{2}
	}, \\
	H_{\rm I, II}(\bs{k}) &=& \frac{h}{4}
	\envelope{pmatrix}{
		+ e^{ + i \bs{k} \cdot \bs{e}_0} (\bs{n} \cdot \bs{e}_0) + e^{ + i \bs{k} \cdot \bs{e}_1} (\bs{n} \cdot \bs{e}_1) &
		+ e^{ + i \bs{k} \cdot \bs{e}_2} (\bs{n} \cdot \bs{e}_2) + e^{ + i \bs{k} \cdot \bs{e}_3} (\bs{n} \cdot \bs{e}_3) \\
		- e^{ + i \bs{k} \cdot \bs{e}_2} (\bs{n} \cdot \bs{e}_2) + e^{ + i \bs{k} \cdot \bs{e}_3} (\bs{n} \cdot \bs{e}_3) &
		+ e^{ + i \bs{k} \cdot \bs{e}_0} (\bs{n} \cdot \bs{e}_0) - e^{ + i \bs{k} \cdot \bs{e}_1} (\bs{n} \cdot \bs{e}_1)
	}, \\
	H_{\rm II, I}(\bs{k}) &=& \frac{h}{4}
	\envelope{pmatrix}{
		+ e^{ - i \bs{k} \cdot \bs{e}_0} (\bs{n} \cdot \bs{e}_0) + e^{ - i \bs{k} \cdot \bs{e}_1} (\bs{n} \cdot \bs{e}_1) &
		- e^{ - i \bs{k} \cdot \bs{e}_2} (\bs{n} \cdot \bs{e}_2) + e^{ - i \bs{k} \cdot \bs{e}_3} (\bs{n} \cdot \bs{e}_3) \\
		+ e^{ - i \bs{k} \cdot \bs{e}_2} (\bs{n} \cdot \bs{e}_2) + e^{ - i \bs{k} \cdot \bs{e}_3} (\bs{n} \cdot \bs{e}_3) &
		+ e^{ - i \bs{k} \cdot \bs{e}_0} (\bs{n} \cdot \bs{e}_0) - e^{ - i \bs{k} \cdot \bs{e}_1} (\bs{n} \cdot \bs{e}_1)
	}.
}
}
Here $\lambda$ is a Lagrange multiplier to ensure the (relaxed) spinon occupation number constraint, 
$ \sum_{\bs{r}} ( |\Phi_{\bs{r}}|^2 - 1 ) = 0$.
Now all $\Phi$-$\Phi$ correlation functions (including the dynamic spin structure factor) 
can be computed from this action.

\section{Fields along other directions and comparison with octupolar U(1)$_0$ QSL}
\label{appC}

Here we include the results of the upper and lower excitations for external fields along $[001]$ 
(Fig.~\ref{fig:excitation_edge_001}) and $[111]$ (Fig.~\ref{fig:excitation_edge_111}) directions.
Enhanced periodicity is also observed in these cases, regardless of the field direction.

\begin{figure}[t]
	\includegraphics[width=.3\textwidth]{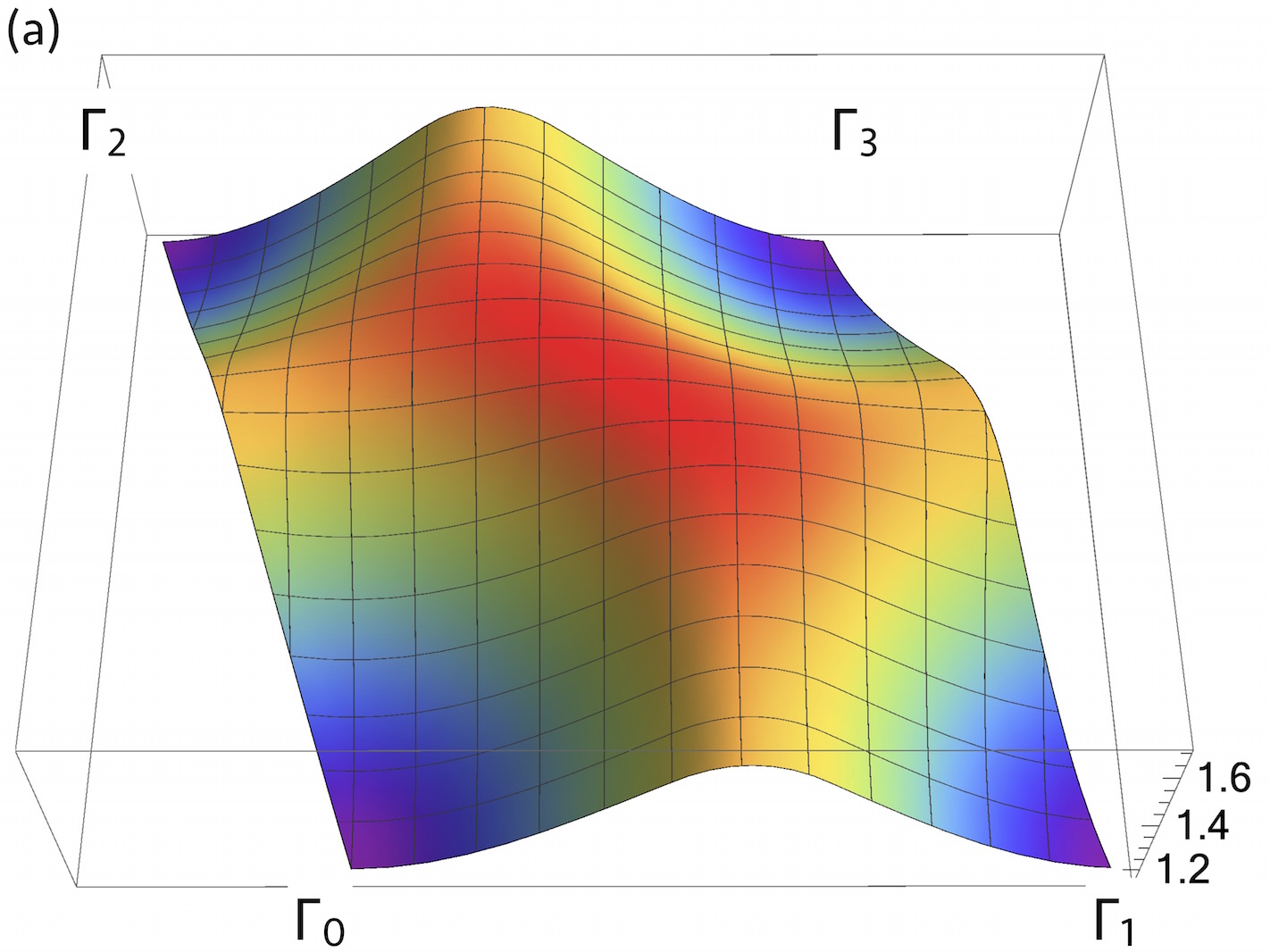}
    \includegraphics[width=.3\textwidth]{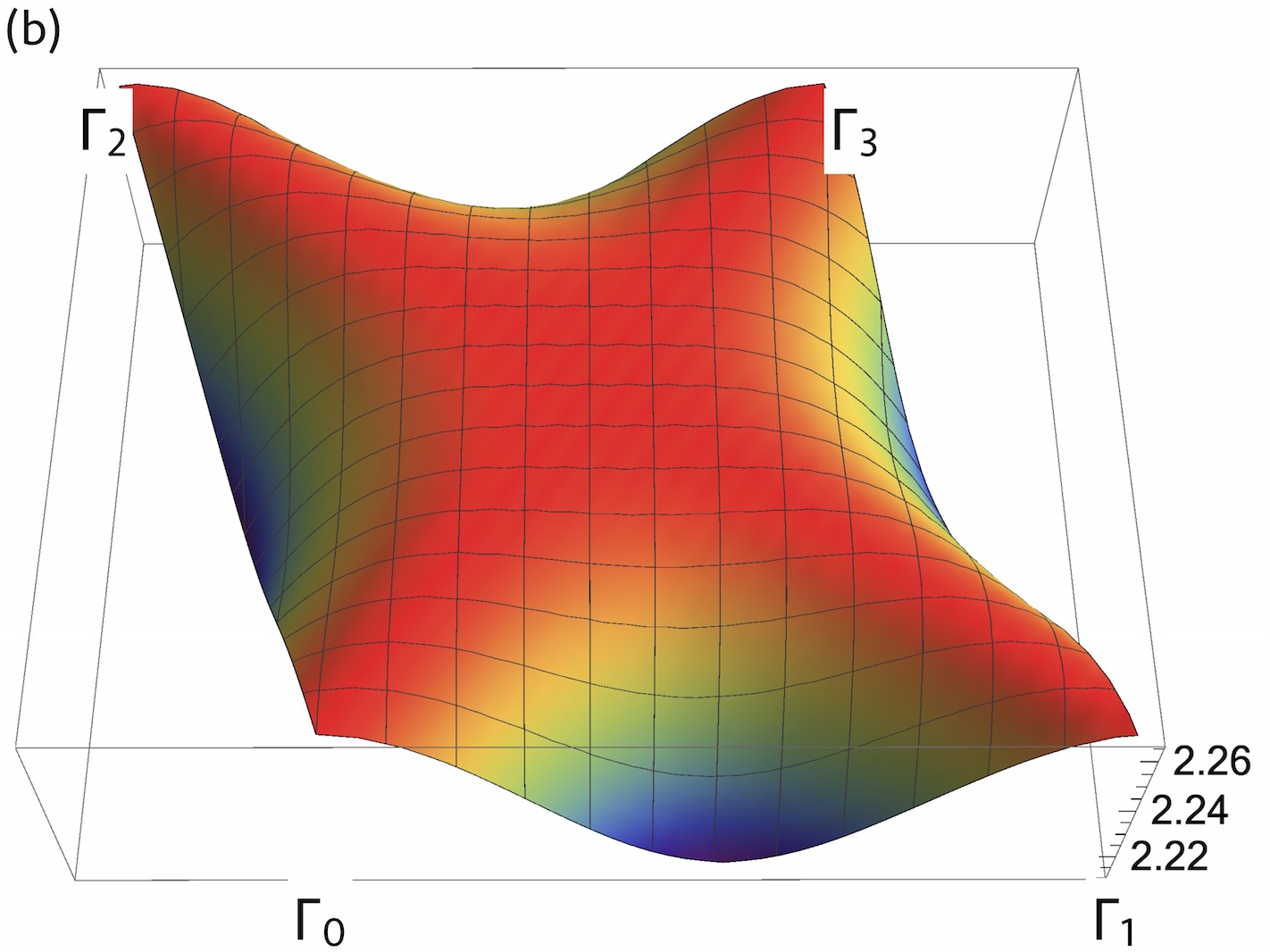}
\caption{(Color online.)
The upper (a) and lower (b) excitation edges of the two-spinon continuum 
in the octupolar $0$-flux U(1) QSL state. We fix $J_{\pm}^y = 0.12 J_y, 
J_{\pm\pm}^y =0$ and $h = 0.0 J_y$. The energy unit is set to be $J_y$. 
}
\label{figx}
\end{figure}

As a comparison, we also present the excitation edge for 
the octupolar $0$-flux U(1) QSL state, where the enhanced 
spectral periodicity is not observed; see Fig.~\ref{figx}.

\bibliography{refs}


\end{document}